\documentclass[11pt]{article}
\usepackage[affil-it]{authblk}
\usepackage{color}
\usepackage{amsmath,amsfonts,amssymb,mathrsfs,amsthm}
\numberwithin{equation}{section}
\usepackage{graphicx}
\usepackage{listings}
\usepackage{enumerate,enumitem}
\usepackage{mathtools}
\usepackage[a4paper, total={6in, 8in}]{geometry}
\usepackage{cite}
\usepackage[colorlinks,linkcolor=blue,citecolor=blue]{hyperref}
\usepackage{empheq}
\usepackage{pdfpages}
\usepackage[titletoc]{appendix}
\usepackage[utf8]{inputenc}
\usepackage{xspace}

\newtheorem{Result}{Result}
\newtheorem{defn}{Definition}

\newcounter{example}[section]

\newcommand{\Figref}[1]{Fig.~\ref{#1}}
\newcommand{\R}{\mathcal{R}}
\newcommand{\C}{\mathcal{C}}

\newcommand{\Sectionref}[1]{Section~\ref{#1}}
\newcommand{\kstar}{\overset{\star}{k}}
\newcommand{\pbar}{\bar{p}}

\newcommand{\Eqref}[1]{Eq.~\eqref{#1}}
\newcommand{\Defref}[1]{Def.~\ref{#1}}
\newcommand{\Eqsref}[1]{Eqs.~\eqref{#1}}

\newcommand{\Resultref}[1]{Result~\ref{#1}}

\newcommand{\DecayO}[2]{O\left( \frac{1}{#1^{#2}} \right)}

\newcommand{\keyword}[1]{\emph{#1}}
\newcommand{\V}{\mathcal{V}}

\title{Asymptotically flat vacuum initial data sets from a modified
  parabolic-hyperbolic formulation of the Einstein vacuum constraint equations}
\author{F.~Beyer\footnote{Email: fbeyer@maths.otago.ac.nz},$\;$
  J.~Frauendiener \footnote{Email: joergf@maths.otago.ac.nz}$\;$ and J.~Ritchie\footnote{Email: jritchie@maths.otago.ac.nz}}
\affil{Department of Mathematics and Statistics, University of Otago, New Zealand.}
\begin{document}
	
	\maketitle
	
	\begin{abstract}
		In this paper we continue earlier investigations \cite{Beyer:2017tu,Beyer:2018HW,csukas2019} of evolutionary formulations of the Einstein vacuum constraint equations originally introduced by R\'{a}cz.  Motivated by the strong evidence from these works that the resulting vacuum initial data sets are generically \emph{not} asymptotically flat we analyse the asymptotics of the solutions of a modified formulation by a combination of analytical and numerical techniques. We conclude that the vacuum initial data sets generated with this new formulation are generically asymptotically flat.
	\end{abstract}

	\section{Introduction}
	The {Einstein vacuum constraint equations} are a subset of the full Einstein field equations (EFE). The triple $(\Sigma,\gamma_{ab},K_{ab})$ of a $3$-dimensional differentiable manifold $\Sigma$, Riemannian metric $\gamma_{a b}$ and a smooth symmetric tensor field $K_{ab}$ on $\Sigma$ is called a \textit{vacuum initial data set} if it satisfies the \textit{Einstein vacuum constraint equations} 
	\begin{align}
	\prescript{(3)}{}{R}-K_{ab}K^{ab}+K^{2}=0,\;\; \nabla_{a}{K^{a}}_{c}-\nabla_{c}K=0,
	\label{VacuumConstraints}
	\end{align} 
	everywhere on $\Sigma$, where $\nabla_{a}$ is the covariant derivative associated with $\gamma_{ab}$, $\prescript{(3)}{}{R}$ is the corresponding Ricci scalar and $K={K^a}_a$ is the mean curvature. For this whole paper we agree that spatial abstract indices $a, b,\dots$ are raised and lowered with the metric $\gamma_{a b}$.
	
	Owing to the work of Choquet-Bruhat and Geroch \cite{FouresBruhat:1952ji,ChoquetBruhat:1969cl} we know that for every solution of the Einstein vacuum constraints there exists a  unique maximal globally hyperbolic development (a solution of the full vacuum EFE). Constructing solutions of the Einstein vacuum constraints is therefore the first crucial step in exploring solutions to the full vacuum EFE. The Einstein vacuum constraints \Eqref{VacuumConstraints} comprise a set of four nonlinear partial differential equations that constrain the twelve independent components of the two tensor fields $\gamma_{ab}$, $K_{ab}$. Solving \Eqref{VacuumConstraints} is therefore an under-determined problem, and to the best of our knowledge, there is no clear physically or geometrically preferred way to construct solutions. 
	
	One of the most successful methods for solving the constraints is the Lichnerowicz-York conformal approach (see \cite{choquet-bruhat2000} and references therein) which allows one to cast the constraints as a set of non-linear elliptic partial differential equations, which can in principle be solved as a boundary value problem. Solving the equations in this way can be challenging. However, there are several well established methods for doing this that have been very successful both from the analytical and the numerical perspective \cite{bartnik2004, Baumgarte:2010vs}. Nevertheless, this approach is not without limitation. For example, mathematical problems have been known to arise when solutions with large mean curvatures are sought  (see \cite{dilts2017,anderson2018a} for an overview and references). Other more physical problems, such as spurious radiation \cite{chu2014,lovelace2009} also occur. Some researchers have therefore sought other methods of solving the constraints \cite{Bishop:1998cb,Matzner:1998hv,Moreno:2002dm,Bishop:2004gb}.
	
	In this work we focus on one such alternative approach, namely, the evolutionary formulations of the vacuum constraints introduced by R\'{a}cz in \cite{Racz:2014kk,Racz:2014dx, Racz:2015bu,Racz:2015gb}. In his work, R\'{a}cz introduced two ways to write the vacuum constraints: as a hyperbolic-algebraic system of PDEs on the one hand, and as a parabolic-hyperbolic system of PDEs on the other hand. In all these cases the constraints are solved as a Cauchy problem similar to earlier work in \cite{bartnik1993, Bishop:1998cb}. 
%
%
	First steps in investigating whether this approach has any advantages over more established methods  have been carried out in \cite{racz2018} for the constraints of the Maxwell equations and in \cite{Nakonieczna:2017vk,doulis2019} for the Einstein vacuum constraint equations.  The main principal disadvantage of R\'{a}cz's approach (in comparison to solving the vacuum constraints as an elliptic boundary value problem) is that it does not directly allow to control the asymptotics of the resulting vacuum initial data sets at spacelike infinity. This is problematic because certain physical quantities such as the total mass or the centre of mass (see for example \cite{Szabados:2009ig,Cerebaum:2016}), to name a few, are only well defined if the data sets satisfy particular asymptotic conditions. With no control over the asymptotics it is therefore possible that the method generates initial data sets that lack a physical interpretation. Exactly this issue has been explored recently in \cite{Beyer:2017tu,Beyer:2018HW,csukas2019}. It was confirmed that generic solutions of these equations are not \emph{asymptotically flat} (this notion is defined in \Sectionref{Sec:Backgrounds} below). This is the case even for small (nonlinear) perturbations of asymptotically flat vacuum initial data sets. Other issues have been observed in \cite{Winicour:2017wr}. 
	
	In \cite{Beyer:2018HW} we proposed an iterative approach to, at least partly, address the asymptotic flatness problem. In contrast to this, 
this paper here provides strong analytical and numerical evidence that a small change of how the free data for R\'acz's parabolic-hyperbolic formulation are specified is sufficient to guarantee asymptotic flatness of the vacuum initial data sets generated by this method. We note that a different, but similarly spirited modification was suggested in \cite{csukas2019}.
As in \cite{Beyer:2017tu,Beyer:2018HW}, we restrict most of our attention to $\Sigma$ being the exterior region of an isolated gravitational source and we mostly assume that $\Sigma$ is foliated by $2$-spheres. This allows us to use the same numerical pseudo-spectral methods developed in \cite{Beyer:2017jw,Beyer:2014bu,Beyer:2016fc,Beyer:2009vw} based on the $\eth$- and the spin-weight formalism. 
We shall discuss that foliations based on topological $2$-spheres imply the restriction that the constraint equations must be solved ``towards spatial infinity'' away from the sources of the gravitational field. We shall label this direction as the \emph{increasing $\rho$-direction} where $\rho$ is the \emph{evolution parameter}.  We remark that our focus here (as well as that of earlier works \cite{Beyer:2017tu,Beyer:2018HW}) on the \emph{asymptotics} of these vacuum initial data sets will be overcome in future work.
In this work here we are indeed not concerned with the properties of the solutions in the strong field regime, e.g.\ of apparent horizons. We also remark that the setup in  \cite{doulis2019}, where foliations in terms of $2$-planes are considered allowing for evolutions ``towards the sources of the gravitation field'', is not well-suited to study the asymptotics because of the necessity of finite boundary conditions on the $2$-planes. 
	
	The paper is outlined as follows: In \Sectionref{Sec:ParabolicHyperbolicConstraints} we briefly summarise the framework of $2+1$-decompositions and introduce Kerr-Schild-like data sets. After a quick summary of R\'{a}cz's \emph{original} parabolic-hyperbolic formulation of the vacuum Einstein constraints in \Sectionref{SubSec:RaczParabolicHyperbolicConstraintsN}, we discuss our new modified version of these equations in \Sectionref{SubSec:AnAlternativeChoiceOfFreeData}. \Sectionref{Sec:Backgrounds} is then devoted to the discussion of the asymptotics; we define the concept of \emph{asymptotic flatness} and what it means for the $2+1$-quantities introduced above. Now \Sectionref{SubSec:Hyperboloidal_Sphereical} yields analytical evidence for our claim that the vacuum initial data sets obtained with our modified parabolic-hyperbolic formulation are better behaved than those with the original formulation in as much as that generic solutions are asymptotically flat. We then support these analytical results by numerics in \Sectionref{Sec:BinaryBlackHoles}.

	\section{Preliminaries}
	\label{Sec:ParabolicHyperbolicConstraints}
	\subsection{The \texorpdfstring{$2+1$}{2+1}-decomposition of initial data sets}	
	\label{SubSec:Racz2Plus1Decomp}	

	We now discuss R\'{a}cz's original parabolic-hyperbolic formulation of the Einstein vacuum constraints. Further details can be found in \cite{Racz:2014kk,Racz:2014dx,Racz:2015gb,Racz:2015bu}. We use the same conventions as in \cite{Beyer:2018HW}.
	
	Consider an arbitrary $3$-dimensional manifold $\Sigma$, Riemannian metric $\gamma_{a b}$ and smooth symmetric tensor field $K_{ab}$; at this stage these are not required to satisfy any equation (such as the vacuum constraints).  As before  the Levi-Civita covariant derivative associated with $\gamma_{a b}$ is labelled $\nabla_{a}$.
	We suppose there exists a smooth function $\rho:\Sigma \rightarrow \mathbb{R}$ whose collection of
level sets $\mathcal{S}_\rho$ forms a foliation of $\Sigma$. This foliation yields a decomposition of $(\Sigma,\gamma_{ab},K_{ab})$, in full analogy to standard $3+1$-decompositions of spacetimes \cite{Alcubierre:Book}, as follows.
	The unit co-normal of any of the $2$-surfaces $\mathcal{S}_\rho$ is 
	\begin{align}
	N_{a}=A\nabla_{a}\rho,
	\label{eq:defNa}
	\end{align}
	where $A>0$ is the \textit{lapse}. 
	The induced first and second fundamental forms are therefore, respectively,
	\begin{align}
	h_{ab}&=\gamma_{ab}-N_{a}N_{b},
	\label{eq:defhdd}\\
	k_{ab}&=-\frac{1}{2}\mathcal{L}_{N}h_{ab}.
	\label{eq:defkdd}
	\end{align}
	We shall label the covariant derivative associated with $h_{ab}$ as $D_{a}$.
	The tensor field
	\begin{align*}
	{h^a}_{b}={\delta^a}_{b}-N^{a}N_{b}
	\end{align*}
	is the map that projects an arbitrary tensor defined at any point $p$  in $\Sigma$ orthogonally to a tensor that is tangent to $S_\rho$ at $p$. 	
	If each index of a tensor field defined on $\Sigma$ contracts to zero with $N_{a}$ or $N^{a}$ at all $p\in\Sigma$, then we call that it \textit{intrinsic (to the foliation of surfaces $S_{\rho}$)}.  Given an arbitrary tensor field on $\Sigma$ we can create an intrinsic tensor field by contracting each
	index with ${h^a}_b$. In fact, any tensor can be uniquely decomposed into its intrinsic and its orthogonal parts, e.g.,
	\begin{align}
	\label{eq:Kdec}
	K_{ab}=\kappa N_{a}N_{b}+N_{a}p_{b}+N_{b}p_{a}+q_{ab},
	\end{align}
	with
	\begin{align}
	\kappa=N^{a}N^{b}K_{ab},\; p_{a}={h^c}_{a}N^{b}K_{cb},\; q_{ab}={h^c}_{a}{h^d}_{b}K_{cd}.
	\label{Eq:Decompose_Kdd}
	\end{align}
	The field $q_{ab}$ is symmetric and can be further decomposed into its trace and trace-free part (with respect to $h_{ab}$) as follows
	\begin{align}
	q_{ab}=Q_{ab}+\frac{1}{2}q h_{ab}, \;\; Q_{ab}h^{ab}=0,
	\label{Eq:Decompose_qdd}
	\end{align}          
	where the relations 
	\begin{align}
	q=h^{ab}q_{ab},\; Q_{ab}h^{ab}=0
	\end{align}
        hold and $Q_{ab}$ is symmetric. 
	
	Now pick an arbitrary vector field  $\rho^a$ such that 
	\begin{align}
	\rho^{a}\nabla_{a}\rho =1.
	\end{align}
	According to \Eqref{eq:defNa}  there must exist a unique intrinsic vector field $B^{a}$, called the \textit{shift}, such that
	\begin{align}
	\rho^a = A N^{a} + B^{a},
	\label{N_rho}
	\end{align}
        where $A$ is the lapse in \Eqref{eq:defNa}.
	Given $\rho^a$, we can write \Eqref{eq:defkdd} as
	\begin{align}
	k_{ab}
	=-A^{-1}\left(\frac{1}{2}\mathcal{L}_{\rho}h_{ab}-D_{\left( a\right.}B_{\left. b\right)} \right)=:A^{-1}\overset{\star}{k}_{ab}.
	\label{eq:defkStar}
	\end{align}
	We also define
	\begin{align}
	\kstar:=h^{ab}\kstar_{ab}.
	\label{eq:defkStar2}
	\end{align}		
	Finally, the Ricci scalar ${}^{(3)}R$ associated with $\gamma_{ab}$ can be written as
	\begin{align}
	{}^{(3)}R=&{}^{(2)}R-\left(A^{-2} \kstar^2 +A^{-2}\kstar_{ab}\kstar^{ab}+2A^{-1}D^{a}D_{a}A-2\left( A^{-1}\mathcal{L}_{N}\overset{\star}{k}-A^{-2}\mathcal{L}_{N}A \right)  \right),
	\end{align}
	where the Ricci scalar associated with the induced metric $h_{ab}$ is called ${}^{(2)}R$.
	The \emph{intrinsic acceleration vector} is 
	\begin{align}
	v_{b}&=N^{a}\nabla_{a}N_{b}= -A^{-1}D_{b}A.
	\label{eq:defvb}
	\end{align}

	\subsection{Kerr-Schild-like data sets}
	\label{SubSec:Kerr_Shild_Like}
	
	In this subsection we introduce data sets (without imposing the constraints yet) of Kerr-Schild form. Such data sets were the basis of our previous work in \cite{Beyer:2017tu,Beyer:2018HW} and we shall continue to use them in particular in \Sectionref{Sec:A_superposition_method_for_generating_binary_black_hole_data}. In this paper now we introduce such data sets as follows.
	\begin{defn}
		\label{Def:KerrSchild}
		A data set $(\Sigma, \gamma_{ab}, K_{ab})$ is called \textit{Kerr-Schild-like} if $\Sigma=\mathbb{R}^3\backslash \overline B$ where $B$ is a  ball in $\mathbb{R}^3$  and there exists a smooth function $V:\Sigma\rightarrow \mathbb{R}$ with $V<1$, a smooth co-vector field $l_{a}$ and a symmetric tensor field $\dot{\gamma}_{ab}$ such that 
		\begin{align}
		\gamma_{ab}=\delta_{ab}-Vl_{a}l_{b},
		\;\;
		K_{ab}=\frac{\sqrt{1-V}}{2}\left(  \nabla_{a}\left( Vl_{b} \right)+\nabla_{b}\left( Vl_{a} \right) -\dot{\gamma}_{ab} \right),
		\label{KerrSchild}
		\end{align}
		where $\delta_{ab}$ is the flat metric on $\Sigma$, $\left(\delta^{-1} \right)^{ab}$ its inverse, and $l_a$ satisfies the condition 
\begin{equation}
\label{eq:KSnormalisation}
\left( \delta^{-1} \right)^{ab}l_{a}l_{b}=1.
\end{equation}
	\end{defn}
An example of a Kerr-Schild like data set is the standard ingoing Kerr-Schild Schwarzschild slice given by $l_a=\nabla_ar$, $V=-2m/r$ and $\dot\gamma_{ab}=0$.
	
	Let us now proceed by providing some useful formulas derived from this definition. For
	\begin{equation}
	\tilde{l}^{a}=(\delta^{-1})^{ab}l_{b},
	\end{equation}
	it follows
	\begin{align}
	\label{eq:lnorm}
	\tilde{l}^{a}l_a&=(\delta^{-1})^{ab}l_a l_b=1,	\\
\gamma^{ab}&=(\delta^{-1})^{ab}+\frac{V}{1-V}\tilde{l}^{a}\tilde{l}^{b},\\	
	\label{eq:normalla}
	l^a&=\frac 1{1-V}\tilde{l}^{a},\quad l^al_a=\frac 1{1-V}.
        \end{align}
	
	\newcommand{\normall}{f}
	Suppose now we have chosen a  smooth function $\rho$ on $\Sigma$ with the properties discussed in \Sectionref{SubSec:Racz2Plus1Decomp} giving rise to a foliation $S$ in terms of level sets $S_\rho$ diffeomorphic to the $2$-sphere. We restrict to the case where $l_a$ is normal to $S_\rho$, i.e.,
	\begin{equation}
	\label{eq:specla}
	l_a=\pm\normall\nabla_a\rho,
	\end{equation}
	with
	\begin{equation}
	\label{eq:lanorm}
	\normall=\frac 1{\sqrt{(\delta^{-1})^{ab}\nabla_a\rho\nabla_b\rho}},
	\end{equation}
        as a consequence of \Eqref{eq:KSnormalisation}.
	From \Eqsref{eq:defNa}, \Eqref{eq:specla} and \Eqref{eq:normalla} we find that
	\begin{equation}
	N_a=\sqrt{1-V}\, l_a,
	\end{equation}
	which means that the lapse defined in \Eqref{eq:defNa} is
	\begin{equation}
	\label{eq:KSA}
	A=\normall \sqrt{1-V}.
	\end{equation}
	It now follows from \Defref{Def:KerrSchild} and \Eqref{eq:defhdd} that
	\begin{equation}
	\label{eq:KShab}
	h_{ab}=\delta_{ab}-l_a l_b.
	\end{equation}
	Since
	\begin{equation}
	\label{eq:KSKab}
	K_{ab}
	=\frac{2-V}{4(1-V)}\left(\nabla_a V N_b+\nabla_b V N_a\right)
	+\frac{V}{2}\left(\nabla_a N_b+\nabla_b N_a\right)
	-\frac{\sqrt{1-V}}{2}\dot\gamma_{ab},
	\end{equation} 
	\Eqref{Eq:Decompose_Kdd} yields
	\begin{align}
	\label{eq:KSkappa}
	\kappa&=\frac{2-V}{2(1-V)^{3/2}} \tilde l^a\nabla_a V 
	-\frac{\sqrt{1-V}}{2}\dot\gamma_{ab}N^aN^b,
	\\
	\label{eq:KSpa}
	p_a&=\frac{2-V}{4(1-V)} D_a V 
	+\frac{V}{2}v_a -\frac{\sqrt{1-V}}{2}\dot\gamma_{cb}{h^{c}}_{a}N^b,
	\\
	\label{eq:KSqab}
	q_{ab}&=-V {k}_{ab}-\frac{\sqrt{1-V}}{2}\dot\gamma_{cd}{h^{c}}_{a}{h^{d}}_{b},
	\end{align}
	where $v_a$ can be calculated from \Eqref{eq:defvb} and $k_{ab}$ from \Eqref{eq:defkStar} once a shift vector field $B^a$, and thereby the vector field $\rho^a=A N^a+B^a$, has been chosen. Notice that we can calculate $B_a$ as
	\begin{equation}
          \label{eq:KSBa}
          B_a=\rho^bh_{ab}.
	\end{equation}
The quantities $q$ and $Q_{ab}$ are given by
	\Eqref{Eq:Decompose_qdd} and $\overset{\star}{k}_{ab}$ and $\kstar$  are obtained from \Eqsref{eq:defkStar} and \eqref{eq:defkStar2}.

\section{Parabolic-hyperbolic formulations of the vacuum constraints}
	\subsection{R\'{a}cz's parabolic-hyperbolic formulation of the vacuum constraints}
	\label{SubSec:RaczParabolicHyperbolicConstraintsN}	
        Given the function $\rho$ and the foliation in terms of $2$-surfaces $S_\rho$ generated by it as in \Sectionref{SubSec:Racz2Plus1Decomp}, the vacuum constraints \Eqref{VacuumConstraints} can now  be decomposed into their normal and intrinsic components, and, according to \cite{Racz:2015gb} yield the following system of equations: 
	\begin{align}
	\overset{\star}{k}\mathcal{L}_{\rho}A+A^2 D^{a}D_a A-\overset{\star}{k}B^a D_{a}A=\,
	&\frac 12 A^3 E
	+\frac 12A F,
	\label{ParabolicEquation_Racz}
	\\
	\mathcal{L}_{\rho}q-B^a D_{a}q-AD_{a}p^{a}-2p^{a}D_{a}A=\,&\overset{\star}{k} {}^{ab}Q_{ab}+\frac{1}{2}q\overset{\star}{k} -\overset{\star}{k}\kappa,
	\label{FinalSystemDiffNorm_Racz}
	\\
	\begin{split}
	\mathcal L_{\rho} p_c-B^a D_{a}p_{c}-\frac{1}{2}AD_{c}q-\kappa  D_c A +{Q^a}_{c}D_{a}A+\frac{1}{2}qD_{c}A=\,&p_a D_b B^a
	-AD_{a} {Q^a}_{c}
	\\
	&+\overset{\star}{k}p_c + AD_{c}\kappa,
	\end{split}
	\label{FinalSystemDiffMom_Racz}
	\end{align}
	\newcommand{\ParabolicHyperbolicR}{\Eqsref{ParabolicEquation_Racz}--\eqref{FinalSystemDiffMom_Racz}\xspace}
	where 
	\begin{align}
	E&={}^{(2)}R+2\kappa q-2p^{a}p_{a}-Q_{ab}Q^{ab}+\frac{1}{2}q^2,
	\\
	F&=2(\partial_{\rho}\overset{\star}{k}-B^a D_{a}\overset{\star}{k})-\overset{\star}{k}_{ab}{\overset{\star}{k}}{}^{ab}-\overset{\star}{k}{}^2.
	\end{align}

	Observe that all quantities here  are smooth \emph{intrinsic} tensor fields. It is clear that while this means that all contractions with $N^a$ or $N_a$ vanish, contractions with $\rho^a$ do not, e.g., 
$p_\rho:=p_a\rho^a=p_aB^a$ as a consequence of \Eqref{N_rho}. However such ``components'' $p_\rho$ do clearly not constitute a further degree of freedom of the field $p_a$ since $p_\rho=p_aB^a$ is fully determined by its ``intrinsic components''. Consistently with this, it is easy to check that 
the equation for $p_\rho$ obtained by contracting \Eqref{FinalSystemDiffMom_Racz} with $\rho^c$ fully decouples from the remaining equations. 
We remark that instead of thinking of each field in the equations above as an intrinsic field on $\Sigma$, we could equivalently think of it as a $1$-parameter family of fields on $\mathbb S^2$ defined by the pull-back along the $\rho$-dependent map $\Phi_\rho: \mathbb S^2\rightarrow\Sigma$, $p\mapsto (\rho,p)$ to $\mathbb S^2$. In the following we shall use abstract indices  $A,B,\ldots$ for such $\rho$-dependent tensor fields  on $\mathbb S^2$. Indeed, all indices $a,b,\ldots$ in the equations above could be replaced by $A,B,\ldots$, and, at the same type, each Lie-derivative along $\rho^a$ by the derivative with respect to parameter $\rho$. All this is well-known for $3+1$-decompositions of spacetimes and is therefore not discussed any further here.

	\ParabolicHyperbolicR suggest to group the various fields introduced above are as follows:
	\begin{description}
		\item[Free data] The fields $B_{a}$, $Q_{ab}$,  $h_{ab}$ and $\kappa$ are considered as {freely specifiable} everywhere
		on $\Sigma$. All of $\kstar$, $D_a$,
		${}^{(2)}R$, $Q_{ab}$ and $F$ (together with all of the index versions of these) as well as all coefficients in \ParabolicHyperbolicR are fully determined by these on
		$\Sigma$.
		\item[Unknowns] The quantities $A$, $q$ and $p_{a}$ are considered as the unknowns of \ParabolicHyperbolicR once free data have been specified.  
	\end{description}
	According to \cite{Racz:2014kk}, it can be shown that given arbitrary smooth \emph{Cauchy data}\footnote{The Cauchy datum for $A$ is assumed to be strictly positive everywhere without further notice.} for $A$, $q$ and $p_a$ on an arbitrary $\rho=\rho_0$-leaf of the $2+1$-decomposition of $\Sigma$, in addition to smooth \emph{free data} everywhere $\Sigma$, the \emph{Cauchy problem} of \ParabolicHyperbolicR  in the \emph{increasing} $\rho$-direction is well-posed, i.e., the equations have a unique smooth solution $A$, $q$ and $p_{a}$ at least in a $\rho\ge\rho_0$-neighbourhood of the initial leaf $S_{\rho_0}$, provided the \keyword{parabolicity condition} holds everywhere on $\Sigma$:
	\begin{equation}
	\label{eq:parabolcond}
	\kstar<0.
	\end{equation}
Clearly, if $\kstar$ is positive instead, then the {Cauchy problem} is well-posed in the \emph{decreasing} $\rho$-direction instead.
	In any case, \ParabolicHyperbolicR is a quasilinear parabolic-hyperbolic system provided \Eqref{eq:parabolcond} holds everywhere.

It is important to remember that since the equation for the lapse is essentially a nonlinear heat equation there is a significant difference between evolving in the ``forward'' and ``backward'' direction -- a notion determined by the sign of $\kstar$ here. The Cauchy problem being well-posed in the forward direction (the increasing $\rho$-direction if $\kstar<0$) means that the solutions are guaranteed to be smooth and well-behaved, while in the backward direction (the decreasing $\rho$-direction if $\kstar<0$) they generically become ``arbitrarily non-smooth after arbitrarily small evolution times''. Certain particular regular solutions may still be found in the backward direction, but the general lack of stability makes the backward problem unsuitable for numerical investigations. We therefore fully focus on the forward Cauchy problem here.

It is interesting to notice that $\kstar$ is fully determined by the free data. The condition \Eqref{eq:parabolcond} can therefore be verified  \emph{prior}  to solving \ParabolicHyperbolicR. From \Eqsref{eq:defkdd}, \eqref{eq:defkStar} and \eqref{eq:defkStar2} we deduce that $\kstar$ has the opposite sign than the mean curvature of the leaves of the foliation. Given \Eqref{eq:defNa} and the assumption that the lapse $A$ is positive, it follows that \emph{the Cauchy problem of \ParabolicHyperbolicR is well-posed in the $\rho$-direction of the increasing area of the leaves of the foliation}. In the particular case  that the foliation is of $2$-sphere topology, as we shall restrict to for most of this paper, we shall align $N^a$ with the outward-pointing direction. Since we expect this to be the direction of increasing area (at least asymptotically), we therefore anticipate \Eqref{eq:parabolcond} to hold and the increasing $\rho$-direction therefore to agree with the outward-pointing direction towards spatial infinity. In this setting all evolutions of \ParabolicHyperbolicR must therefore be performed in the increasing $\rho$-direction.

	\subsection{Modified parabolic-hyperbolic formulation of the vacuum constraints}
	\label{SubSec:AnAlternativeChoiceOfFreeData}
	
The system \ParabolicHyperbolicR  has been used in several works among which are 
\cite{Beyer:2018HW,csukas2019,doulis2019,winicourRacz2018,Nakonieczna:2017vk}.
	The particular choice of how to split the fields into free data and unknowns is however not the only possibility. Motivated by previous studies \cite{Beyer:2018HW,csukas2019}, which indicate an instability of these equations in the asymptotically flat setting, we now propose a small modification. The main result of our paper is that we can provide evidence that this instability observed for \ParabolicHyperbolicR is resolved by this modification. 

Recall that $\kappa$ is one of the free data in the  formulation introduced in \Sectionref{SubSec:RaczParabolicHyperbolicConstraintsN} while $q$ is one of the unknowns. Here now we propose to introduce a new free data field $\R$
and then set
	\begin{equation}
	\label{eq:alternativedatadec}
	\kappa=\R q
	\end{equation}
	where $q$ continues to be an \emph{unknown}. 
	The  equations resulting from this are obtained from \ParabolicHyperbolicR by replacing all instances of $\kappa$ with $\R q$:
	\begin{align}
	\overset{\star}{k}\mathcal{L}_{\rho}A+A^2 D^{a}D_a A-\overset{\star}{k}B^a D_{a}A=\,
	&\frac 12 A^3 E
	+\frac 12A F,
	\label{ParabolicEquation}
	\\
	\mathcal{L}_{\rho}q-B^a D_{a}q-AD_{a}p^{a}-2p^{a}D_{a}A=\,&\overset{\star}{k} {}^{ab}Q_{ab}+\frac{1}{2}q\overset{\star}{k} -\overset{\star}{k}\R q,
	\label{FinalSystemDiffNorm}
	\end{align}
	\begin{align}
	\begin{split}
	\mathcal L_{\rho} p_c-B^a D_{a}p_{c}-A\left( \frac{1}{2}+\R \right)D_{c}q=\,&p_a D_b B^a
	-AD_{a} {Q^a}_{c} + q\R  D_c A -{Q^a}_{c}D_{a}A
	\\
	&+\overset{\star}{k}p_c + AqD_{c}\R-\frac{1}{2}qD_{c}A,
	\end{split}
	\label{FinalSystemDiffMom}
	\end{align}
	\newcommand{\ParabolicHyperbolicA}{\Eqsref{ParabolicEquation}--\eqref{FinalSystemDiffMom}\xspace}
	where, $F$ takes the same form as before and $E$ becomes
	\begin{align}
	E&={}^{(2)}R-2p^{a}p_{a}-Q_{ab}Q^{ab}+\left( 2\R + \frac{1}{2} \right) q^2.
	\end{align} 
	We shall refer to these equations as the \emph{modified parabolic-hyperbolic system} while  \ParabolicHyperbolicR shall often be labeled as the \emph{original parabolic-hyperbolic system}.
	
	First we observe that this modification has changed the principal part of the system. 
	It turns out that \ParabolicHyperbolicA is still parabolic-hyperbolic. First, the principal part of \Eqref{ParabolicEquation} is unchanged (and is therefore parabolic provided the same parabolicity condition \Eqref{eq:parabolcond} as before holds), and, second, the subsystem \Eqsref{FinalSystemDiffNorm} -- \eqref{FinalSystemDiffMom} is symmetrisable hyperbolic with symmetriser 
	\begin{align}
	\left(
	\begin{array}{cc}
	\frac{1}{2}+ \R & 0 \\
	0 & h^{ce}
	\end{array}
	\right) 
	\end{align}
	provided 
	\begin{align}
	\frac{1}{2}+\R>0,
	\label{HyperbolicityCondition}
	\end{align}
	where $h^{ce}$ is the intrinsic inverse of $h_{ab}$.
	We  refer to \Eqref{HyperbolicityCondition} as the \textit{hyperbolicity condition}.
	This now suggests the following choice: 
	\begin{description}
		\item[Free data:] The fields $B_{a}$, $Q_{ab}$, $h_{ab}$ and $\R$ are {free data} everywhere on $\Sigma$. 
		\item[Unknowns:] The fields $A$, $q$ and $p_{a}$ are the unknowns. 
	\end{description}
	It follows that for arbitrary free data, for which both the parabolicity condition \Eqref{eq:parabolcond} and the hyperbolicity condition \Eqref{HyperbolicityCondition} hold, \ParabolicHyperbolicA is a quasilinear parabolic-hyperbolic system and
	the \emph{Cauchy problem} in the increasing $\rho$-direction is therefore well-posed (at least locally). Both conditions \Eqsref{eq:parabolcond} and \eqref{HyperbolicityCondition} are conditions on the free data as before.
	We remark that our hyperbolicity condition here  should not be confused with the hyperbolicity condition found by R\'{a}cz in his so-called algebraic-hyperbolic formulation \cite{Racz:2015gb}. 

It is not obvious why \ParabolicHyperbolicA should be ``any better'' than \ParabolicHyperbolicR. The rest of the paper is about exactly this issue.

	\section{Asymptotics and radial expansions of data sets (without imposing the vacuum constraints yet)}
	\label{Sec:Backgrounds}

	As in \cite{Beyer:2017tu,Beyer:2018HW} we restrict now to the case $\Sigma=\mathbb{R}^3\backslash \overline B$ where $B$ is an arbitrary fixed ball in $\mathbb{R}^3$ in all of what follows. Moreover, we assume  that the level sets of $\rho$ are diffeomorphic to $2$-spheres. This implies that we can assume that
	\[\Sigma=(\rho_{-},\infty)\times\mathbb S^2\]
	for some $\rho_{-}>0$ and we write the points in $\Sigma$ as $(\rho,p)$ with $\rho\in (\rho_{-},\infty)$ and $p\in\mathbb S^2$. Observe carefully that we often use the same symbol $\rho$ for the real parameter $\rho\in (\rho_{-},\infty)$ and for the \emph{function} $\rho$ defined by $(\rho,p)\mapsto \rho$ used for the $2+1$-decomposition.
	Consider now the manifold $\Sigma=(\rho_{-},\infty)\times
	\mathbb S^2$ for some $\rho_{-}>0$ as before. An initial data set (not necessarily a solution of the vacuum constraints\footnote{Initial data sets that are \emph{solutions} of the vacuum constraints are discussed in the  sections following this one.}) is equivalently specified by  a Riemannian metric $\gamma_{ab}$ and smooth symmetric tensor field $K_{ab}$ on $\Sigma$, or, by
	the fields $(A,\kappa,q,p_{a},B_{a},Q_{ab},h_{ab})$ on $\Sigma$ as in \Sectionref{SubSec:Racz2Plus1Decomp}. We shall often speak of $(A,\kappa,q,p_{a},B_{a},Q_{ab},h_{ab})$ as \emph{the $2+1$-fields associated with $(\gamma_{ab},K_{ab})$}, or, equivalently of $(\gamma_{ab},K_{ab})$ as the \emph{initial data set associated with the $2+1$ quantities} $(A,\kappa,q,p_{a},B_{a},Q_{ab},h_{ab})$. 
	
	Let us now introduce some more notation and further structure. Given any $\rho\in (\rho_{-},\infty)$, let $\Phi_\rho: \mathbb S^2\rightarrow\Sigma$ be the map $p\mapsto (\rho,p)$ introduced earlier. Recalling the index conventions before,  we let $(\Omega^{-1})^{AB}$ be the contravariant round unit metric on $\mathbb S^2$. Sometimes it is useful to use standard polar coordinates $(\vartheta,\varphi)$ on $\mathbb S^2$ in terms of which the components of $(\Omega^{-1})^{AB}$ take the form of the matrix $\mathrm{diag}(1,\sin^{-2}\vartheta)$. 	
	Given now an arbitrary smooth intrinsic tensor field ${T_{a\ldots b}}$ on $\Sigma$, let $T_{A\ldots B}$ be the ($\rho$-dependent) pull-back to $\mathbb S^2$ as discussed before. 
	We  then define the $\rho$-dependent norm  
	\begin{align}
	|{T_{a\ldots b}}|^2:=T_{A'\ldots B'}T_{A\ldots B}(\Omega^{-1})^{AA'}\cdots (\Omega^{-1})^{BB'}.
	\end{align}
	Notice that this is a norm only for \emph{intrinsic} tensor fields on $\Sigma$. Given this
	we write ${T_{a\ldots b}}=\DecayO{\rho}{k}$ provided there is a uniform constant $C$ such that $|{T_{a\ldots b}}|\le C \rho^{-k}$ sufficiently close to $\rho=\infty$. We say that ${T_{a\ldots b}}$ has an \emph{asymptotic radial expansion of order $k$} (near $\rho=\infty$) provided
	\begin{align}
	{T_{a\ldots b}}=\sum_{i=0}^{k-1} {{T^{(i)}}_{a\ldots b}}\rho^{-i}+\DecayO{\rho}{k},
	\end{align}
	where the coefficients ${{T^{(i)}}_{a\ldots b}}$ are smooth intrinsic tensor fields on $\Sigma$ which do \emph{not} depend on $\rho$, i.e., $\mathcal L_\rho {{T^{(i)}}_{a\ldots b}}=0$. If $T_{a\ldots b}=O(1)$ then we say $T_{a\ldots b}$ as an \textit{asymptotic radial expansion of order 0}. In order to simplify the notation, we sometimes shall use these notions of the norm and the $O$-symbol  for \emph{general} tensor fields on $\Sigma$ even when they are not intrinsic. In this case observe that this norm and this $O$-symbol are ``completely blind'' to all ``transversal components'' of the tensor field. 
	
	For the following it is also useful to define $\Omega_{ab}$ as the tensor field on $\Sigma$ with the property $\rho^a\Omega_{ab}=0$ whose pull-back along the map $\Phi_\rho$ above equals the covariant round unit metric on the $2$-sphere for each $\rho$, i.e.,  the inverse of $(\Omega^{-1})^{AB}$. Notice carefully that $\Omega_{ab}$ defined this way is \emph{not} intrinsic to the foliation (unless the shift vector field $B^a$ vanishes). Its components with respect to adapted coordinates $(\rho,\vartheta,\varphi)$ on $\Sigma$ as introduced before correspond to the matrix $\mathrm{diag}(0,1,\sin^2\vartheta)$.
	
	In all of what follows we shall assume without further notice that $\rho_{-}$ is sufficiently large so that all $2+1$-quantities  are well-defined. 	
	Recall that asymptotically flat data sets have been studied by us before in \cite{Beyer:2017tu,Beyer:2018HW} where we have we used the same definitions originally from \cite{Dain:2001cd}.
	\begin{defn}
		\label{AsympFlatDef}     
		The triple $(\Sigma,\gamma_{ab},K_{ab})$ with $\Sigma=\mathbb{R}^3\backslash \overline B$ where $B$ is a  ball in $\mathbb{R}^3$
		is called an asymptotically flat initial data set provided
		there exist coordinates $\{ x^i \}$ on $\Sigma$ such that the components of
		$\gamma_{a b}$ and $K_{ab}$ with respect to these coordinates satisfy, respectively, 
		\begin{align}
		\gamma_{ij}=\left( 1 + \frac{2M}{R} \right)\delta_{ij}+\DecayO{R}{2},
		\;\;\,
		K_{ij}= \DecayO{R}{2},
		\end{align}
		in the limit 
		\begin{align}
		R=\left(\sum_{a=1}^{3} (x^i)^2 \right)^{1/2} \rightarrow \infty,
		\label{FlatLimit}
		\end{align} 
		where $\delta_{ij}=\mathrm{diag}(1,1,1)$. The quantity $M\in\mathbb R$
		is called the \textit{ADM mass}.
	\end{defn}
	Asymptotic flatness therefore implies conditions on the asymptotics of $2+1$ quantities associated with an initial data set $(\Sigma,\gamma_{a b},K_{ab})$; see also \cite{Beyer:2017tu,Beyer:2018HW,csukas2019}.
	\begin{Result}[\textbf{Asymptotically flat data sets}]
		\label{Thm_Flt}
		A data set $(\Sigma,\gamma_{ab}, K_{ab})$ is asymptotically flat with ADM mass $M$ provided all corresponding $2+1$-fields have the following asymptotic radial expansions:
		\begin{enumerate} 
			\item The expansion of $A$ is of order $2$ with $A^{(0)}=1$ and $A^{(1)}=M=\mathrm{const}$.
			\item The expansion of $B_a$ is of order $1$ with
			$B^{(0)}_a=0$.
			\item The expansion of $h_{ab}$ is of the form $\rho^{-2} h_{ab}=\Omega_{ab}+O(\rho^{-2})$.
			\item The expansion of $q$ is of order $2$ with $q^{(0)}=q^{(1)}=0$.
			\item The expansion of $p_a$ is of order $1$ with $p_a^{(0)}=0$.
			\item The expansion of $Q_{ab}$ is $Q_{ab}=O(1)$.
			\item The expansion of $\kappa$ is of order $2$ with $\kappa^{(0)}=\kappa^{(1)}=0$.
		\end{enumerate}		
	\end{Result}
	\begin{proof}		
		As before we assume that $\Sigma=(\rho_{-},\infty)\times\mathbb S^2$ with radial parameter $\rho$. For the following it is useful to introduce coordinates $(\rho,\vartheta,\varphi)$ on $\Sigma$ where $(\vartheta,\varphi)$ are standard polar coordinates on each leaf diffeomorphic to $\mathbb S^2$. As mentioned before the components of $\Omega_{ab}$ with respect to these coordinates take the form $\mathrm{diag}(0,1,\sin^2\vartheta)$. Under the assumptions above, the components of $\gamma_{ab}$ with respect to these coordinates are
		\[\gamma_{\alpha\beta}=\mathrm{diag}\left(1+\frac{2 A^{(1)}}\rho,\rho^2,\rho^2\sin^2\vartheta\right)+
		\begin{pmatrix}
		\DecayO{\rho}{2} & \DecayO{\rho}{} & \DecayO{\rho}{}\\
		\DecayO{\rho}{} & O(1) & O(1)\\
		\DecayO{\rho}{} & O(1) & O(1)
		\end{pmatrix},
		\]
		where the $O$-symbol for each component here is interpreted as that for \textit{scalar functions} on $\mathbb R^3$.
		With respect to the new radial coordinate
		\[R=\rho-A^{(1)},\]
		the components of $\gamma_{ab}$ are therefore
		\[\gamma_{\alpha'\beta'}=\omega^2\mathrm{diag}\left(1,R^2,R^2\sin^2\vartheta\right)+
		\begin{pmatrix}
		\DecayO{R}{2} & \DecayO{R}{} & \DecayO{R}{}\\
		\DecayO{R}{} & O(1) & O(1)\\
		\DecayO{R}{} & O(1) & O(1)
		\end{pmatrix},
		\]
		where {$\omega^2=1+\frac{2A^{(1)}}{R}$}. Transforming the polar coordinates $(R,\vartheta,\varphi)$ to Cartesian coordinates in the standard way, we finally obtain
		\[\gamma_{ij}=\omega^2\mathrm{diag}\left(1,1,1\right)+
		\begin{pmatrix}
		\DecayO{R}{2} & \DecayO{R}{2} & \DecayO{R}{2}\\
		\DecayO{R}{2} & \DecayO{R}{2} & \DecayO{R}{2}\\
		\DecayO{R}{2} & \DecayO{R}{2} & \DecayO{R}{2}
		\end{pmatrix},
		\]
		as required for asymptotic flatness. We can therefore identify $A^{(1)}$ with the quantity $M$. The same arguments applied to $K_{ab}$  
		yield that the condition for asymptotic flatness is satisfied provided $\kappa^{(0)}=\kappa^{(1)}=0$, $p_a^{(0)}=0$ and $q_{ab}=O(1)$ (which is equivalent to assumptions 4 and 6).
	\end{proof}
		
Given an arbitrary initial data set (not necessarily solving the constraints), then we can show that\footnote{Without going into technical details we assume here that the $O$-symbol does not only control the fields themselves as discussed before, but also sufficiently many of their derivatives in the natural way.}
$\kstar=-2/\rho+O(1)$, and therefore $\kstar<0$ for sufficiently large $\rho$; cf.~\Eqref{eq:parabolcond}. Since general asymptotically flat data sets of the form in \Resultref{Thm_Flt} imply that $\R=O(1)$, they can only be used as \emph{backgrounds} for solving the modified system \ParabolicHyperbolicA, if we impose additional conditions to ensure \Eqref{HyperbolicityCondition}. We discuss this issue below.

	Here now we return briefly to Kerr-Schild-like data sets introduced in \Sectionref{SubSec:Kerr_Shild_Like}. To this end we introduce an arbitrary smooth function $r$ on $\Sigma$ with the property that 
	\begin{equation}
	\label{eq:KSr}
	r=\rho+O(\rho^{-1});
	\end{equation}
	notice carefully that we demand that no $O(1)$-term is present in this expansion. In terms of this function $r$, we assume that the flat metric $\delta_{ab}$ in \Defref{Def:KerrSchild} takes the form
	\begin{equation}
          \label{eq:KSdelta}
	\delta_{ab}=\nabla_a r \nabla_b r+r^2\Omega_{ab}
	\end{equation}
	where $\Omega_{ab}$ was introduced above. Given this	
	it is straightforward to show that the function $\normall$ in \Eqsref{eq:specla} and \eqref{eq:lanorm} is
	$\normall=1+O(\rho^{-2})$, that $\rho^{-2}h_{ab}=\Omega_{ab}+O(\rho^{-2})$ as a consequence of \Eqref{eq:KShab} and that $B_a=O(\rho^{-1})$ from \Eqref{eq:KSBa}. It follows from \Resultref{Thm_Flt} and the formulas in \Sectionref{SubSec:Kerr_Shild_Like} that the Kerr-Schild-like data set is asymptotically flat provided $V$ has an asymptotic radial expansion of order $2$ where $V^{(0)}=0$ and $V^{(1)}=const$. In this case the ADM mass is $M=-V^{(1)}/2$.

	\section{Vacuum initial data sets obtained by the modified parabolic-hyperbolic system}
	\label{sec:sphsymm}
	
	\subsection{The spherically symmetric case}
	\label{SubSec:Hyperboloidal_Sphereical}	

        In this section we analyse the asymptotics of vacuum initial sets obtained as solutions of \ParabolicHyperbolicA. Recall that \ParabolicHyperbolicR have been analysed in \cite{Beyer:2018HW,csukas2019}. We  present evidence that all the instabilities regarding asymptotic flatness, which were found for the original system, are resolved by this modification.

The general idea here and in the following is to pick a \emph{background} initial data set (in general not a solution of the constraints) which is asymptotically flat according to \Resultref{Thm_Flt} in a first step. From this background data set, we then read off the free data for solving \ParabolicHyperbolicA in a second step.
We start this subsection with the simpler spherically symmetric case in which \ParabolicHyperbolicA reduces to a system of ordinary differential equations. 
To this end we consider  backgrounds in Kerr-Schild-like form as in \Sectionref{SubSec:Kerr_Shild_Like} with \Eqsref{eq:KSr} -- \eqref{eq:KSdelta}.  We impose spherical symmetry by requiring that $V$ only depends on $\rho$ and that $r=\rho$. We also choose $\dot{\gamma}_{ab}=0$. The  $2+1$-quantities defined by this are
	\begin{align}
	Q_{ab}=0,\;\; h_{ab}=\rho^{2}\Omega_{ab},\;\; B_{a}=0,\;\; \R=\frac{(2-V)\rho}{4\left( 1- V \right)}\frac{\partial_\rho V}{V},
	\label{GeneralFreeData}
	\end{align}
	and
	\begin{align}
	\begin{split}
	q=\frac{2V}{\rho\sqrt{1-V}},\;\;  p_{a}=0,\;\; A =\sqrt{1-V}.
	\label{GeneralDecomp}
	\end{split}
	\end{align}
	In order to ensure that $\R$ is a smooth quantity, we assume that $\partial_{\rho}V/V$ is well-defined and finite for all $\rho>0$.
	
	We use \Eqref{GeneralFreeData} now as a background to determine the free data for the modified parabolic-hyperbolic system \ParabolicHyperbolicA. Since $q$, $p_a$ and $A$ are supposed to be found as solutions of the equations we therefore ignore \Eqref{GeneralDecomp}.
	In order to appeal to spherically symmetry, we  look for solutions under the restriction $p_a =0$ and  where the unknowns $A$ and $q$ only depend on $\rho$. With this, \ParabolicHyperbolicA take the form
	\begin{align}
	\partial_{\rho}q &= -\frac{2}{\rho}\left( \frac{1}{2} - \R \right)q,
	\label{FlatSpecialPDE_R_q}
	\\
	\partial_{\rho}A &= -\frac{\rho}{4}\left( \frac{2}{\rho^2} + \left( 2\R + \frac{1}{2}\right)q^{2}  \right)A^{3} +\frac{1}{2\rho}A.
	\label{FlatSpecialPDE_R_A}
	\end{align}
	It is surprising\footnote{This is not possible for the original system; see \cite{Beyer:2018HW}.} that for any function $V$ which satisfies the previous restrictions, we can write down the general solution explicitly as
	\begin{align}
	q&=\frac{2\C\, V}{\rho\sqrt{1- V }},
	\label{FlatGeneralSol_R_q}\\
	A&=\sqrt{\frac{( 1 - V )\rho}{\rho -2m- ( \rho - 2m )V + \rho\, \C^2 V^2 }},
	\label{FlatGeneralSol_R_A}
	\end{align}
	where $m,\C\in\mathbb{R}$ are free constants. It is interesting to notice that this only agrees with \Eqref{GeneralDecomp} if $\C=1$ and $V=-2m/\rho$. 
	Irrespective of the choice of $V$, the Hawking mass \cite{hawking1968} of each surface $S_\rho$ of the resulting vacuum initial data set turns out to be 
	\begin{align}
	m_{H}=m,
	\end{align}
and is therefore independent of $\rho$.
	
	Since we study vacuum solutions in somewhat more detail in the next subsection,  let us now consider the following specific choice of the function $V$
	\begin{align}
	V=-\frac{\V}{\rho},
	\label{V_Flt_Sphere}
	\end{align}
	for an arbitrary constant $\V\in\mathbb R$. From the discussion at the end of \Sectionref{Sec:Backgrounds}, the background data set above is therefore asymptotically flat with mass $\V/2$.
	With this choice the solutions \Eqref{FlatGeneralSol_R_q}--\eqref{FlatGeneralSol_R_A} have the following asymptotic expansions 
	\begin{align}
	A=1+\frac{m}{\rho} + \DecayO{\rho}{2},\;\; q =-\frac{2\C \V}{\rho^2}+\DecayO{\rho}{3},\;\; \kappa= \R q = \frac{\C \V}{\rho^2}+\DecayO{\rho}{3}.
	\label{My_Flat_Expand}
	\end{align}
	It is a consequence of \Resultref{Thm_Flt} that the resulting vacuum initial data set is therefore 
	\emph{asymptotically flat with ADM mass $m\in\mathbb R$ irrespective of the choice of $\V>0$ and $\C\in\mathbb R$}. In contrast to our findings in \cite{Beyer:2018HW} for the original system, this demonstrates that the modified parabolic-hyperbolic system ``performs significantly better'' and in a far more stable manner in the asymptotically flat setting. It is interesting that the background mass $\V/2$ and the ADM mass $m$ of the resulting vacuum data set are generally distinct.
	
	\subsection{Asymptotic radial expansions of vacuum initial data sets (without symmetries)}
	\label{Sec:Solving_the_constraints_on_asymptotically_spherical_backgrounds}

	In this section we use asymptotic expansions to study the asymptotics of vacuum initial data sets obtained by the modified system \ParabolicHyperbolicA{} for a large class of backgrounds without imposing symmetries. Assuming certain asymptotic radial expansions are valid and the free data satisfy appropriate assumptions, we demonstrate that the solutions of the constraints are \emph{always asymptotically flat} in consistency with our findings in the spherically symmetric case in \Sectionref{Sec:Solving_the_constraints_on_asymptotically_spherical_backgrounds}.
In the section following this one, we then support the strong assumptions which we are required to make here by numerical computations. 	
	We focus on the modified system \ParabolicHyperbolicA. We refer to \cite{Beyer:2018HW} for a corresponding result for the original system \ParabolicHyperbolicR which demonstrates that general solutions of the original system are \emph{not} asymptotically flat.
	\begin{Result}
		\label{Result:AsymAnal_Flt_A}
		Let $\Sigma=(\rho_-,\infty)\times\mathbb S^2$ for some $\rho_->0$. Consider arbitrary smooth free data fields $\R$, $B_a$, $Q_{ab}$ and $h_{ab}$ on $\Sigma$ with the properties:
		\begin{enumerate}
			\item The scalar function $\R$ has an asymptotic radial expansion of order $2$ such that $\R^{(0)}=-1/2$ and $\R^{(1)}$ is a strictly positive function.
			\item The intrinsic covector field $B_{a}$ has  an asymptotic radial expansion of order $1$ with $B^{(0)}_{a}=0$.
			\item The symmetric tracefree intrinsic tensor field $Q_{ab}$  has an asymptotic radial expansion of order $2$ with $Q^{(0)}_{ab}=Q^{(1)}_{ab}=0$.
			\item The symmetric intrinsic tensor field $h_{ab}$  has an asymptotic radial expansion of the form $\rho^{-2}h_{ab}=\Omega_{ab} + O\left( \rho^{-2} \right)$.
		\end{enumerate}
Then the parabolicity and the hyperbolicity conditions, see \Eqsref{eq:parabolcond} and \eqref{HyperbolicityCondition}, hold for sufficiently large $\rho$, and, for any solution $A$, $q$, $p_a$ of 
		the modified parabolic-hyperbolic system \ParabolicHyperbolicA with the properties
		\begin{enumerate}
			\item $A$ is strictly positive and has an asymptotic radial expansion of order $2$,
			\item $q$ has an asymptotic radial expansion of order $2$,
			\item $p_{a}$ is an intrinsic co-vector field with an asymptotic radial expansion of order $2$,
		\end{enumerate}
                we find
                \[q^{(0)}=q^{(1)}=0,\quad p_a^{(0)}=p_b^{(1)}=0,\quad A^{(0)}=1,\quad A^{(1)}=\mathrm{const}.\]
                The vacuum initial data set corresponding to the $2+1$-quantities $(A,q,p_{a},\R,B_{a},Q_{ab},h_{ab})$ is therefore 
		asymptotically flat with ADM mass $A^{(1)}$.
	\end{Result}

The conditions for the free data fields are compatible with \Resultref{Thm_Flt}. Observe, however, that the restriction for $Q_{ab}$ and $\R$ are in fact stronger than the ones required by \Resultref{Thm_Flt}. The additional condition on $\R$ ensures that \Eqref{HyperbolicityCondition} holds in addition to \Eqref{eq:parabolcond}. It is a non-trivial outcome of the analysis that \Resultref{Result:AsymAnal_Flt_A} would in general not hold if
$\R^{(0)}\neq -1/2$. 

	\begin{proof}[Proof of \Resultref{Result:AsymAnal_Flt_A}]
We have discussed before
		\begin{align}
		\kstar = -\frac{2}{\rho} +\DecayO{\rho}{2},
		\end{align}
as a consequence of the hypothesis and that therefore \Eqref{eq:parabolcond} holds for sufficiently large $\rho$.
		We also find that 
		\begin{align}
		\R + \frac{1}{2} = \frac{\R^{(1)}}{\rho} + \DecayO{\rho}{2}
		\label{Eq:Hyperbolicity_Result_Flat}
		\end{align}
                and that the assumption
		 $\R^{(1)}>0$ therefore implies \Eqref{HyperbolicityCondition} for sufficiently large $\rho$ as well. 		\ParabolicHyperbolicA is therefore parabolic-hyperbolic asymptotically. Now we attempt to solve 
		 \ParabolicHyperbolicA order by order in $\rho$. The two leading orders of \Eqref{FinalSystemDiffNorm} immediately imply $q^{(0)}=q^{(1)}=0$. Given this, the leading order of \Eqref{ParabolicEquation} yields the equation
		\begin{align}
		\hat{\Delta}A^{(0)}=-\frac{1-\left(A^{(0)}\right)^2 }{A^{(0)}}=: F[A^{(0)}],
		\label{eq:A0eq}
		\end{align}
		where $\hat{\Delta}$ is the Laplace operator associated with the round $2$-sphere metric $\Omega_{AB}$. It is clear that $A^{0)}=0$ cannot be a solution and we rule out all negative solutions by assumption. One positive solution is $A^{(0)}=1$, in fact, this is the only smooth strictly positive solution: Suppose there were two different smooth strictly positive solutions $A^{(0)}$ and $\tilde A^{(0)}$ of \Eqref{eq:A0eq}. Then a standard integration by parts argument implies
		\begin{equation}
		\label{eqLA0eq2}
		-\|\hat D (A^{(0)}-\tilde A^{(0)})\|^{2} =\left<A^{(0)}-\tilde A^{(0)},F[A^{(0)}]-F[\tilde A^{(0)}]\right>,
		\end{equation}
		where the norm and the scalar product here are the standard $L^2$-norm and $L^2$-scalar product on the $2$-sphere with respect to $\Omega_{ab}$. One can easily check that
		\begin{align}
		F[A^{(0)}]-F[\tilde A^{(0)}]= \frac{A^{(0)}\tilde{A}^{(0)} + 1 }{ A^{(0)}\tilde{A}^{(0)} }\left( A^{(0)} - \tilde{A}^{(0)} \right).
		\end{align}
		Since the fraction on the right-hand side is strictly positive if $A^{(0)}$ and $\tilde A^{(0)}$ are strictly positive, the right-hand side of \Eqref{eqLA0eq2} is therefore non-negative. Since the left-hand side however is non-positive, the implies that $A^{(0)}$ and $\tilde A^{(0)}$ can differ at most by a constant. However, one can easily check that $A^{(0)}=1$ is the only positive constant solution. 		
		Given this, the two leading orders of \Eqref{FinalSystemDiffMom} imply that $p_a^{(0)}=p_a^{(1)}=0$. Finally, we look at the third order ($\rho^{-3}$)-term of \Eqref{FinalSystemDiffMom} to get 
		\begin{align}
		\hat{\Delta}A^{(1)}=0,
		\end{align}
		from which we conclude that $A^{(1)}$ is an arbitrary constant. \Resultref{Thm_Flt} now implies that these solutions are asymptotically flat and that $A^{(1)}$ is the ADM mass.
	\end{proof}

	\section{Numerical investigations}
	\label{Sec:BinaryBlackHoles}        

	\subsection{Black hole background data sets}
	\label{Sec:A_superposition_method_for_generating_binary_black_hole_data}

        Our analytical results in \Sectionref{sec:sphsymm} suggest that general solutions of the \emph{modified} parabolic-hyperbolic system \ParabolicHyperbolicA are asymptotically flat provided the free data satisfy certain asymptotic conditions, in contrast to solutions of the original system \ParabolicHyperbolicA; see \cite{Beyer:2018HW}. In this section now we support these results by numerical calculations.

\begin{figure}[t]
		\centering
		\includegraphics[width=0.8\linewidth]{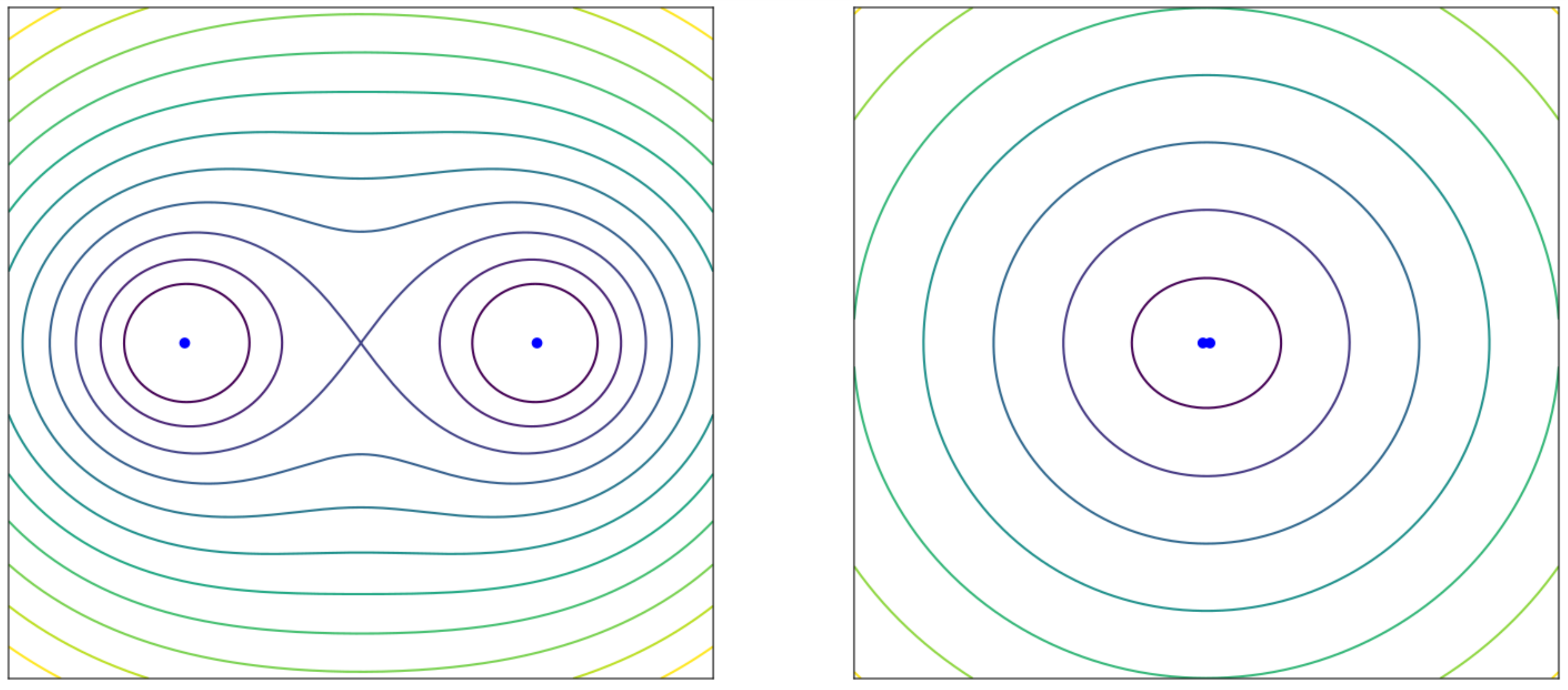}
		\caption{Contour plots of the function $\rho(x,y,z)$ defined by
                  \Eqsref{eq:defuBBH} -- \eqref{eq:defuBBH2} for
                  $M_{+}=M_{-}>0$ and $Z>0$. The left plot shows
                  $\rho$ close to the two black hole positions, while the right figure shows contours for large distances. Both plots restrict to the $x$-$z$-plane. It is evident that the contours become round $2$-spheres in the limit of large distances.}
		\label{fig:contourshyp}
	\end{figure}

In \cite{Beyer:2018HW} we introduced a framework to construct, in principle, multiple black hole background data sets which then provide the free data to solve the constraint equations in a next step. Here we give a short summary of our procedure which is based on  the formalism presented in \Sectionref{SubSec:Kerr_Shild_Like}.	
	Inspired by the ideas presented in \cite{Bishop:1998cb} we
        imagine to have $N$ black hole-like bodies at coordinate positions $(x_i,y_i,z_i)$ with masses $M_i$ for $i=1,\ldots,N$. 
        Using $(x,y,z)$ to label Cartesian coordinates on $\Sigma$ and setting
	\begin{align}
          \label{eq:defuBBH}
	u(x,y,z)=\sum_{i=0}^{N}\frac{M_{i}}{r_{i}(x,y,z)},
	\end{align}
	where 
	\begin{align}
	r_{i}(x,y,z)=\sqrt{(x-x_i)^2 +(y-y_i)^2 +(z-z_i)^2},
	\end{align}
	we define the function $\rho$ as
	\begin{align}
          \label{eq:defuBBH2}
	\rho(x,y,z)=\frac{\sum_{i=0}^{N} M_i}{u(x,y,z)}.
	\end{align}
	Here we restrict to the binary case $N=2$ and write $M_1=M_+$, $M_2=M_-$, $r_1=r_+$ and $r_2=r_-$ where 
	\begin{align}
	r_{\pm}=( 0,0,\pm Z_{\pm}),
	\end{align}  
	for constants $Z_\pm \ge 0$. In contrast to \cite{Beyer:2018HW} we now impose a ``centre of mass condition'' (the reason for this is given below)
	\begin{align}
          \label{eq:centerofmass}
	Z_+ M_{+}-Z_{-}M_{-}=0,
	\end{align}        	
 and therefore choose
\begin{equation}
  \label{eq:Zchoice}
  Z_{-}=Z,\quad Z_{+}=\frac{M_{-}}{M_+}Z,
\end{equation}
for some $Z\ge 0$.
        Notice that $M_-=0$ together with $Z=0$ yields the case of a
        \emph{single black hole}.
        \Figref{fig:contourshyp} shows examples of contour plots of the function $\rho$.         
        It is clear from \Figref{fig:contourshyp} that there is a critical value of $\rho$ where the surfaces undergo a topology change (a \emph{bifurcation}). For $\rho<\rho_{crit}$, each contour is the union of two disconnected $2$-spheres where
	\begin{align}
	\rho_{crit}= Z\frac{\left( M_+ + M_- \right)^2}{M_+\left( \sqrt{M_+} + \sqrt{M_-} \right)^{2}}.
	\label{Eq:BirfurificationPoint}
	\end{align}
        However,
	each $\rho=\text{const}$-surface is diffeomorphic to a \emph{single} $2$-sphere if $\rho>\rho_{crit}$. \Eqref{Eq:BirfurificationPoint} holds under the assumption that $M_+>0$ and $M_-,Z\ge 0$. In all of what follows we restrict to this latter exterior regime of $\mathbb{R}^3$ where the collection of $\rho=\text{const}$-surfaces give rise to a foliation. We emphasise that there certainly exist regular foliations with $2$-sphere topology other than the one given by \Eqsref{eq:defuBBH} -- \eqref{eq:defuBBH2} which extend arbitrarily close to the black holes in a regular fashion as well as to the asymptotic regime. Exploring the wide range of possibilities here will be important in future studies of \emph{both} the strong field regime close to the black holes and the asymptotic far field regime. In this work here we restrict completely on the latter for which the exterior foliation given by 
 \Eqsref{eq:defuBBH} -- \eqref{eq:defuBBH2} and \Eqref{Eq:BirfurificationPoint} is sufficient.
	
	Finally,
given all this, we pick 
\begin{align}
  V = -2u,\;\;\; \dot{\gamma}_{ab}=0,
\end{align}
and
 \begin{equation}
   \label{eq:choiceofr}
   r = \sqrt{x^2 + y^2 + z^2},
 \end{equation}
 as in \cite{Beyer:2018HW} and then obtain an initial data set (not necessarily a solution of the constraints) in the Kerr-Schild form using \Sectionref{SubSec:Kerr_Shild_Like} together with \Eqref{eq:KSdelta}. We find that the resulting data set agrees with the Schwarzschild Kerr-Schild data set with mass $M_+$ in the single black hole case $M_-=Z=0$. Moreover, we see easily that \Eqref{eq:KSr} holds as a consequence of the centre of mass condition \Eqref{eq:centerofmass}.

	We can now show by direct calculations that  for any $M_{+}$, $M_{-}$ and $Z$ as above, the hypothesis of \Resultref{Result:AsymAnal_Flt_A} for $\R$, $B_a$, $Q_{ab}$ and $h_{ab}$ are satisfied at least for all sufficiently large $\rho$. The hypothesis about the unknown fields $A$, $q$ and $p_a$ can, however, as a matter of principle, not be verified a-priori. The main purpose of the following numerical experiments is to provide evidence that the conclusions of \Resultref{Result:AsymAnal_Flt_A}, namely that the resulting vacuum initial data sets are always asymptotically flat, hold nevertheless.

\subsection{Numerical setup}
Given a background data set in \Sectionref{Sec:A_superposition_method_for_generating_binary_black_hole_data}, the next task is to numerically solve the Cauchy problem of \ParabolicHyperbolicA with free data determined by this background. We explain below that we use two different ways to specify Cauchy data in the two following numerical examples. As discussed in more detail in \cite{Beyer:2018HW}, while the background data sets are given in Cartesian coordinates $(x,y,z)$ on $\Sigma$, or, equivalently in corresponding spherical coordinates $(r,\theta,\phi)$ using \Eqref{eq:choiceofr}, the evolutions of \ParabolicHyperbolicA must be performed in adapted $2+1$-coordinates $(\rho,\vartheta,\varphi)$ where $\rho$ is given by \Eqsref{eq:defuBBH} -- \eqref{eq:defuBBH2} and where $(\vartheta,\varphi)$ are intrinsic polar coordinates on each $\rho=\mathrm{const}$-surface diffeomorphic to $\mathbb S^2$. As in \cite{Beyer:2018HW} we choose
\[\vartheta=\theta,\quad\varphi=\phi.\]
This completely fixes the coordinate transformation between the two coordinate systems $(r,\theta,\phi)$ and $(\rho,\vartheta,\varphi)$ on $\Sigma$.

Since the exterior region is foliated by $2$-spheres, we can apply the \emph{spin-weight formalism} following  \cite{Penrose:1984tf,Beyer:2015bv,Beyer:2014bu,Beyer:2016fc,Beyer:2017jw,Beyer:2017tu}. A brief summary is given in \Sectionref{Sec:SWSHstuff} in the appendix. We  express the covariant derivative operator $D_a$ (defined with respect to the intrinsic metric $h_{ab}$)  in terms of the covariant operator $\hat D_a$ defined with respect to the round unit-sphere metric $\Omega_{ab}$; recall that $D_a-\hat D_a$ can be expressed by some smooth intrinsic tensor field. Using \Sectionref{Sec:SWSHstuff}, we can then express the covariant derivative operator $\hat D_a$ in terms of the $\eth$- and $\eth'$-operators \cite{Penrose:1984tf}. Once all of this has been completed for all terms in \ParabolicHyperbolicA, each of these equation and each term end up with a consistent well-defined spin-weight. Most importantly, however, all terms are explicitly regular: Standard polar coordinate issues at the poles of the $2$-sphere disappear when all quantities are  expanded in terms of \keyword{spin-weighted spherical harmonics} and \Eqsref{eq:eths} and \eqref{eq:eths2} are used to calculate the intrinsic derivatives. From the numerical point of view this gives rise to a (pseudo)-spectral scheme. We can therefore largely reuse the code presented in \cite{Beyer:2018HW} subject to two minor changes: (1) the definition of $\rho$ now allows that $Z_{+}\ne Z_{-}$ in agreement with \Eqref{eq:Zchoice}, and, (2) all instances of $\kappa$ in the equations are now replaced with $\R q$ in agreement with our \emph{modification} which leads to \ParabolicHyperbolicA. These two changes do not significantly affect our numerical methods. 
Once the appropriate changes were made to the code, convergence tests (analogous to the ones presented in \cite{Beyer:2018HW}) were carried out and the appropriate behaviour was observed. All of the following simulations were carried out using the adaptive SciPy ODE solver \textit{\textbf{odeint}}\footnote{See \url{https://docs.scipy.org/doc/scipy/reference/generated/scipy.integrate.odeint.html}.}. 

Notice that the background data sets constructed in \Sectionref{Sec:A_superposition_method_for_generating_binary_black_hole_data} are axially symmetric and hence there is no dependence on the angular coordinate $\varphi=\phi$. Motivated by this we restrict to numerical solutions of \ParabolicHyperbolicA with that same symmetry in all of what follows. We can therefore restrict to the axisymmetric case of the spin-weight formalism in \Sectionref{Sec:SWSHstuff}.

	\subsection{Axisymmetric perturbations of single Schwarzschild black hole initial data}
	\label{SubSec:Perturbing_spherically_symmetric_initial_data_sets}
	
In this section now we use the background data set given in  \Sectionref{Sec:A_superposition_method_for_generating_binary_black_hole_data} for $M_{+}=1$ and $M_{-}=Z=0$ (the ``single black hole case''). The free data  for \ParabolicHyperbolicA are therefore given by \Eqref{GeneralFreeData} with $V =- 1/\rho$. It follows from \Sectionref{sec:sphsymm} that 
  \begin{align}
	\mathring{q} =-\frac{2}{\rho^{3/2}\sqrt{\rho+1}}, \;\;\mathring{A}=\sqrt{1+\frac{1}{\rho}},\;\;\mathring{p}_{a}=0,
    \label{FlatBackgroundSolution}
  \end{align}
  is then a particular solution of \ParabolicHyperbolicA representing  single Schwarzschild black hole initial data of unit mass (in spherical symmetry).
The point is now to generate axisymmetric (non-linear) \emph{perturbations}  of  this solution  by solving \ParabolicHyperbolicA with the same free data, but with the following \emph{perturbed Cauchy data} imposed at\footnote{For the single black-hole case the foliation does not bifurcate, see \Eqref{Eq:BirfurificationPoint}, and so all values $\rho_{ 0 }>0$ are allowed.} $\rho_{0}=3$:
\begin{align}
  \label{eq:perturbedCD}
	\left. q \right|_{\rho=\rho_{0}}= \left. \mathring{q} \right|_{\rho=\rho_{0}}+\varepsilon\sin\left( \theta \right),\;\; \left. A \right|_{\rho=\rho_{0}}= \left. \mathring{A} \right|_{\rho=\rho_{0}}+\varepsilon\sin\left( \theta \right),\;\; \left. p_a \right|_{\rho=\rho_{0}}= 0,
	\end{align}
	for some freely specifiable constant $\varepsilon\in\mathbb R$. For small values of $\varepsilon$, we can interpret the resulting vacuum initial data sets as perturbations of single Schwarzschild black hole initial data. 

Given these background data and Cauchy data, we then numerically solve \ParabolicHyperbolicA. Using the formalism in \Sectionref{Sec:SWSHstuff} these equations take the form
	\begin{align}
	\partial_{\rho}A &= -\frac{\rho}{4}\left( \frac{2}{\rho^2}\left( 1-2p\pbar \right) + \left( 2\R + \frac{1}{2}\right) q^2  \right)A^3 +\frac{1}{2 \rho}\left( 1 + A\eth\left( \bar{\eth}\left( A \right) \right) \right)A,
	\label{GeneralPDE_dA_2}
	\\
	\partial_{\rho}q &= \frac{1}{\sqrt{2}\rho^2}\left( \bar{\eth}\left( p\right) + \eth\left( \pbar \right)  \right)A -\frac{2}{\rho}\left( \frac{1}{2} - \R \right)q+\frac{2}{\rho^2 \sqrt{2}}\left( p\, \bar{\eth}\left(A \right)+\pbar\, \eth\left(A\right) \right),
	\label{GeneralPDE_dq_2}
	\\
	\partial_{\rho}p &= A\left( \frac{1}{2}+\R \right)\eth\left(q\right) -\frac{2}{\rho}p +\frac{1}{\sqrt{2}}\left( \R -\frac{1}{2} \right) q\, \eth\left( A \right) ,
	\label{GeneralPDE_p2_2}
	\\
	\partial_{\rho}\pbar &= A\left( \frac{1}{2}+\R \right)\bar{\eth}\left(q\right) -\frac{2}{\rho}\pbar +\frac{1}{\sqrt{2}}\left( \R -\frac{1}{2} \right)q\, \bar{\eth}\left( A \right),
	\label{GeneralPDE_p1_2}
	\end{align}
	where
	\begin{align}
	\label{Eq:pDef}
	p=\frac 1{\sqrt 2} p_a \left(\partial_{\vartheta}^a-{\text{i}}\, {\csc\theta}\, \partial_\varphi^a\right),\quad
	\bar p=\frac 1{\sqrt 2} p_a \left(\partial_{\vartheta}^a+{\text{i}}\, {\csc\theta}\, \partial_\varphi^a\right),
	\end{align}
        and, see \Eqref{GeneralFreeData},
        \[\R=-\frac 12+\frac 1{4(1+\rho)}.\]
        The quantities $A$ and $q$ have spin-weight zero, while $p$ and $\pbar$ have spin-weight $1$ and $-1$, respectively.
        For this particular symmetry (and the particular representation of the underlying bundle) we can assume that
        \[p=\pbar.\]

 In order to present our numerical calculations now and use them to check the predictions from \Resultref{Result:AsymAnal_Flt_A} we consider the $\sup$-norm over $\mathbb S^2$ defined, for any smooth scalar function $\mathcal F(\rho,\vartheta)$ (such as $A$ and $q$ above), as
        \begin{align}
	\| \mathcal{F} \|(\rho)=\max_{\vartheta\in [0,\pi]}| \mathcal{F}(\rho,\vartheta) |.
	\label{Eq:SupNorm}
	\end{align} 
        For $p_a$, this norm is defined as
        \begin{align}
	\| p \|(\rho)=\max_{\vartheta\in [0,\pi]}\sqrt{\Omega^{ab}p_a (\rho,\vartheta) p_b (\rho,\vartheta)}
          =\max_{\vartheta\in [0,\pi]}\sqrt{p (\rho,\vartheta)\pbar (\rho,\vartheta)}.
	\end{align}         
  
\begin{figure}[t]
		\centering
		\includegraphics[width=1.0\linewidth]{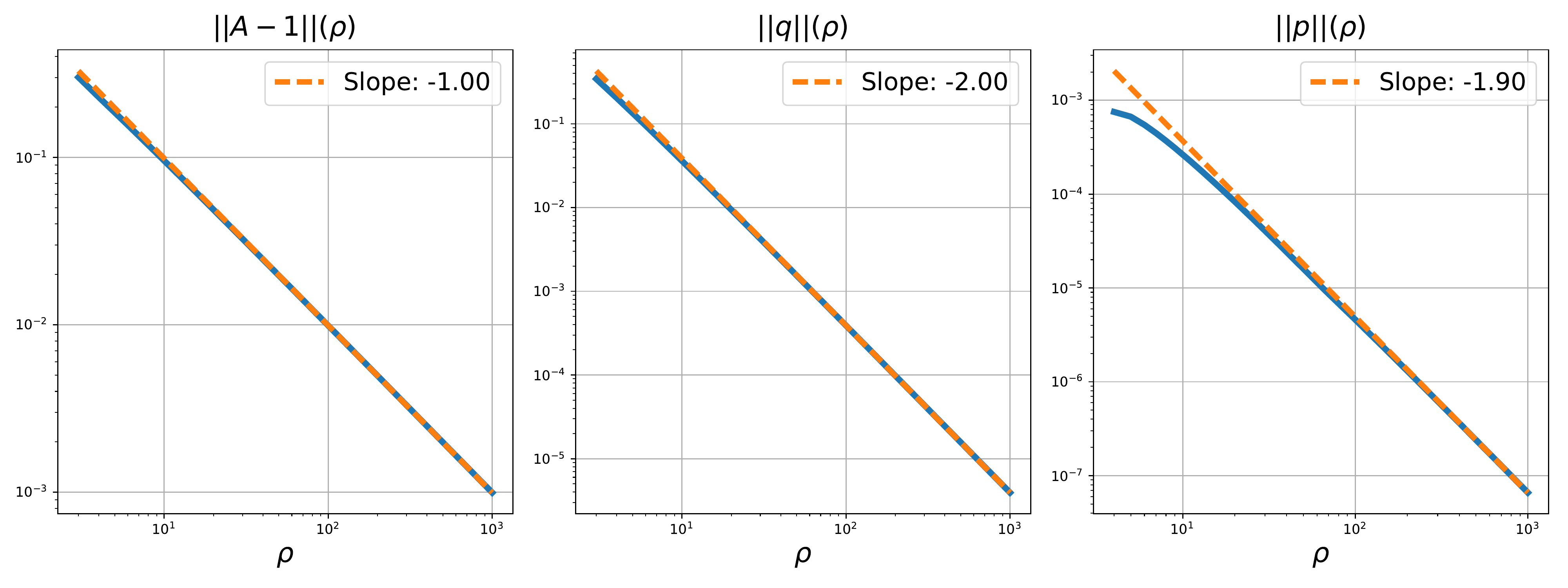}
		\caption{Decay plots of the numerical solution for the ``single black hole case'' obtained with $\epsilon = 10^{-2}$, $N=11$ and a numerical error tolerance of $10^{-12}$.}
		\label{fig:decaynormssbh}
	\end{figure}

	In a first instance, we expect the following behaviour
	\begin{align}
	\| A -1 \|(\rho)= \DecayO{\rho}{},\quad \| q \|(\rho) = \DecayO{\rho}{2},\quad \| p\|(\rho) = \DecayO{\rho}{2}
	\end{align} 
      for all of the solutions above according to \Resultref{Result:AsymAnal_Flt_A}.      	
	\Figref{fig:decaynormssbh} shows that the numerical solutions are indeed consistent with this. The particular numerical solution shown there was  produced with $\varepsilon=10^{-2}$, an absolute and relative error tolerance for the adaptive ODE solver of $10^{-12}$, and for $N=11$, where $N$ is the number of spatial points in the $\vartheta$-direction. We have repeated the same numerical experiments with smaller values of $\varepsilon$ as well and found the same qualitative behaviour in agreement with \Resultref{Result:AsymAnal_Flt_A}. 
\begin{figure}[t]
  \begin{minipage}{0.49\linewidth}
	\centering
	\includegraphics[width=0.85\linewidth]{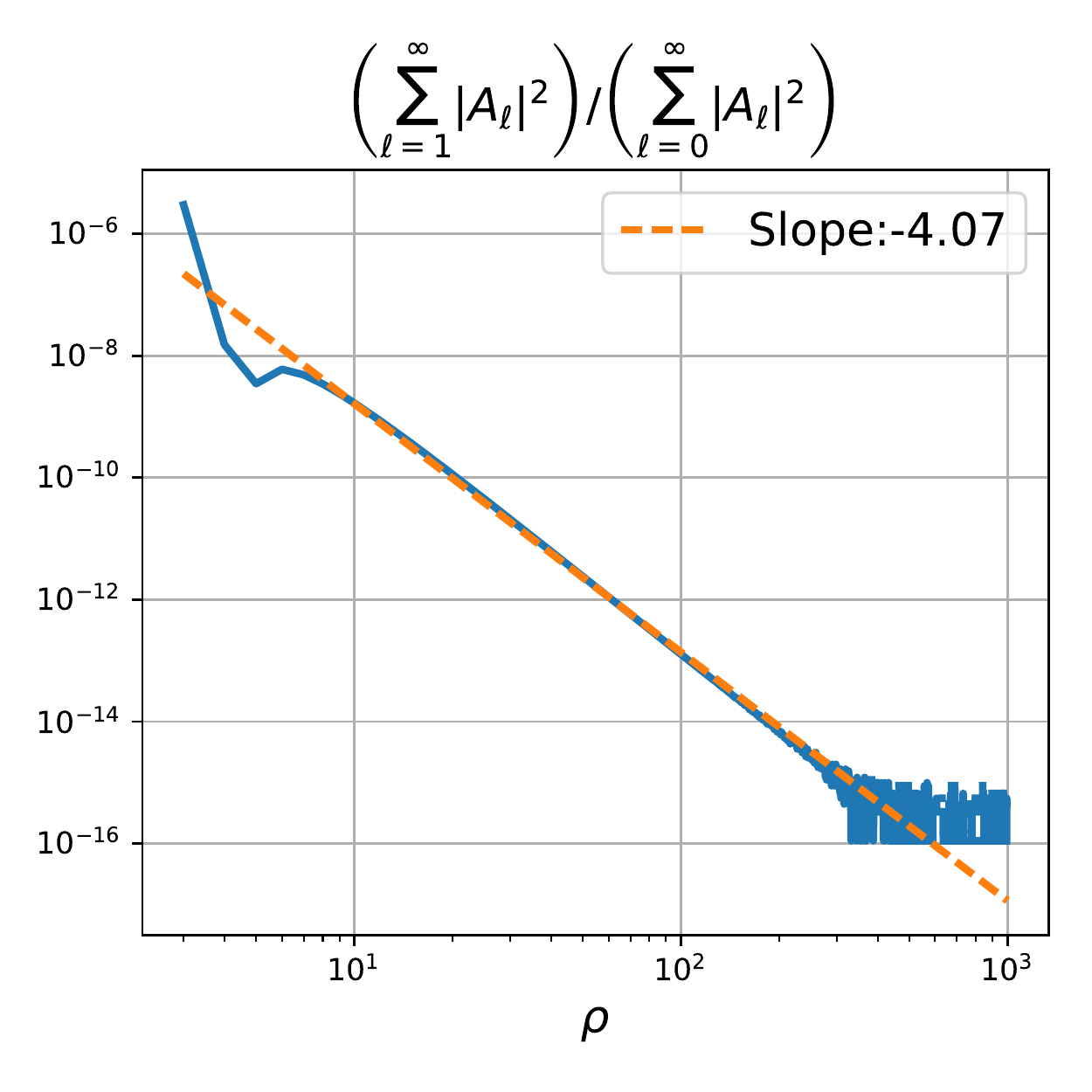}
	\caption{Mode decay plot of the numerical solution for the ``single black hole case''  obtained with the same parameters as \Figref{fig:decaynormssbh}.}
	\label{fig:fftdecaysbh}
      \end{minipage}%
      \hfill%
      \begin{minipage}{0.49\linewidth}
	\centering
      \includegraphics[width=\linewidth]{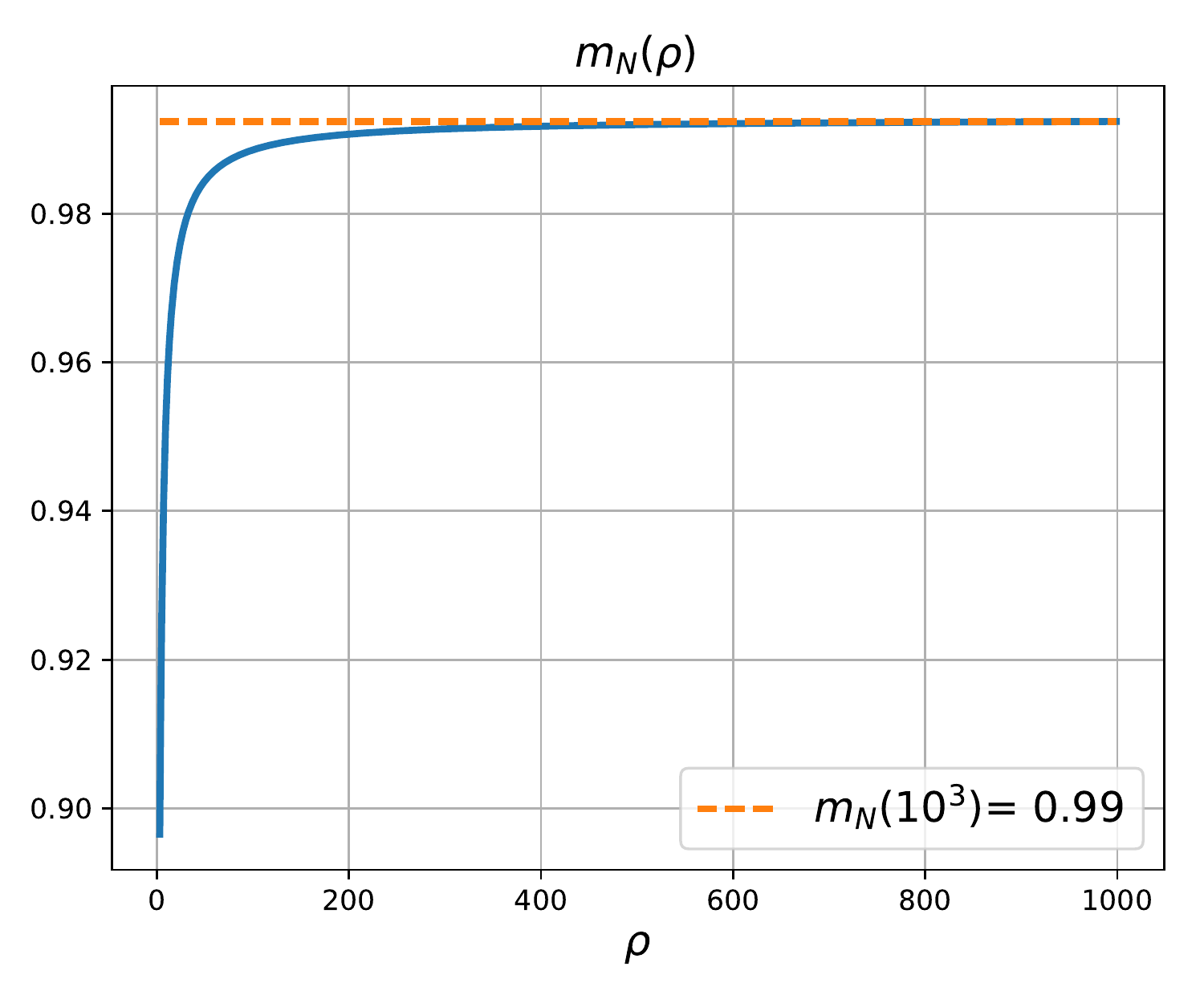}
	\caption{Estimate the ADM mass for the ``single black hole case''  obtained with the same parameters as \Figref{fig:decaynormssbh}.}
	\label{fig:massplotsbh}
      \end{minipage}
    \end{figure}

However, in order to be demonstrate full consistency with \Resultref{Result:AsymAnal_Flt_A} we must show that
\begin{align}
	A=1 + \frac{{A}^{(1)}}{\rho} + \DecayO{\rho}{2},
	\end{align}
for a \emph{constant} $A^{(1)}$ (which then represents the ADM mass). We proceed as follows to numerically support the claim that this is indeed true. If the first two orders of $A$ are constant with respect to $\vartheta$, then the quantity, see \Eqsref{ec:mean_value_s2} and \eqref{eq:L2},
\begin{align}
1-\frac{4\pi | \underline{A}(\rho) |^2}{ \| A(\rho) \|_{L^{2}(\mathbb{S}^2)} }=\frac{\| A(\rho) \|_{L^{2}(\mathbb{S}^2)}-4\pi | \underline{A}(\rho) |^2}{\| A(\rho) \|_{L^{2}(\mathbb{S}^2)}}=\frac{\sum_{\ell=1}^{\infty}|A_\ell (\rho)|^2 }{ \sum_{\ell=0}^{\infty}|A_\ell (\rho)|^2 }
\end{align}
must decay like $O(\rho^{-4})$. In \Figref{fig:fftdecaysbh} we see that this is indeed the case for $\varepsilon = 10^{-2}$.

Let us now discuss how we numerically calculate the ADM mass. In accordance with \Resultref{Result:AsymAnal_Flt_A}, we have 
\begin{align}
\underline{A}(\rho)=1 + \frac{\underline{A}^{(1)}}{\rho} + \DecayO{\rho}{2},
\end{align}
see \Eqref{eq:average}.
Since $\underline{A}^{(1)}={A}^{(1)}$ follows from the above, we therefore find
\begin{align}
A^{(1)} = \rho \left( \underline{A}(\rho) - 1 \right) + \DecayO{\rho}{}.
\end{align} 
This suggests that we define	
\begin{align}
m_{N}(\rho) = \rho \left( \underline{A}(\rho) - 1 \right),
\label{Eq:ADM_Numerical}
\end{align}
as a numerical estimate for the ADM mass $m_{ADM}$. In particular, we get 
\begin{align}
\label{eq:admmassconvergence}
m_{N}(\rho)=m_{ADM}+\DecayO{\rho}{},
\end{align}
as confirmed by \Figref{fig:massplotsbh}.
Given all this it becomes clear that the numerical estimate for the mass $m_{N}$ becomes better as $\rho$ becomes larger. We find, however, that the numerical errors in numerically solving the constraints become significant if we go further than $\rho\sim 10^{3}$. It is natural then to wonder how good the approximation $m_{ADM}=m_N(10^3)$ is. For this we consider the quantity 
	\begin{align}
	\mathcal{E}_{A}[m_{ADM}]=| m_{N}(2\rho)-m_{N}(\rho)|,
	\end{align}
	which is calculated for $\rho=10^3$ as a measure of the absolute error. For our example case, with $\epsilon=10^{-2}$, we find 
	\begin{align}
	m_{ADM} = 0.9942, \quad \mathcal{E}_A[m_{ADM}]=2.34 \times 10^{-6}.
	\end{align}
	Notice that the relative error is of order $\sim 10^{-6}$. As was mentioned above, this is likely due to the error associated with measuring $m_{ADM}$ at a \emph{finite} value of $\rho$. However, due to the errors generated by numerically solving the constraints for very large values of $\rho$, we need to accept whatever error we have at that point in the measurement of the mass.
	
	\subsection{Binary black hole-like initial data sets}
        \label{SubSec:BBH_initial_data_sets}
        In this subsection we repeat essentially the same numerical experiments as before with two changes: (1), the background data set is now determined with parameters $M_{+}=M_{-}=1/2$ and $Z=1$ (an ``equal mass binary black hole case''), and (2),  instead of the ``perturbed'' Cauchy data as in \Eqref{eq:perturbedCD}, we now choose the values obtained from the background data set at $\rho_{ 0 }=3$. For this particular case \Eqref{Eq:BirfurificationPoint} gives that the bifurcation occurs at $\rho_{crit}=1$. 
	
        \begin{figure}[t]
	 	\centering
	 	\includegraphics[width=1.0\linewidth]{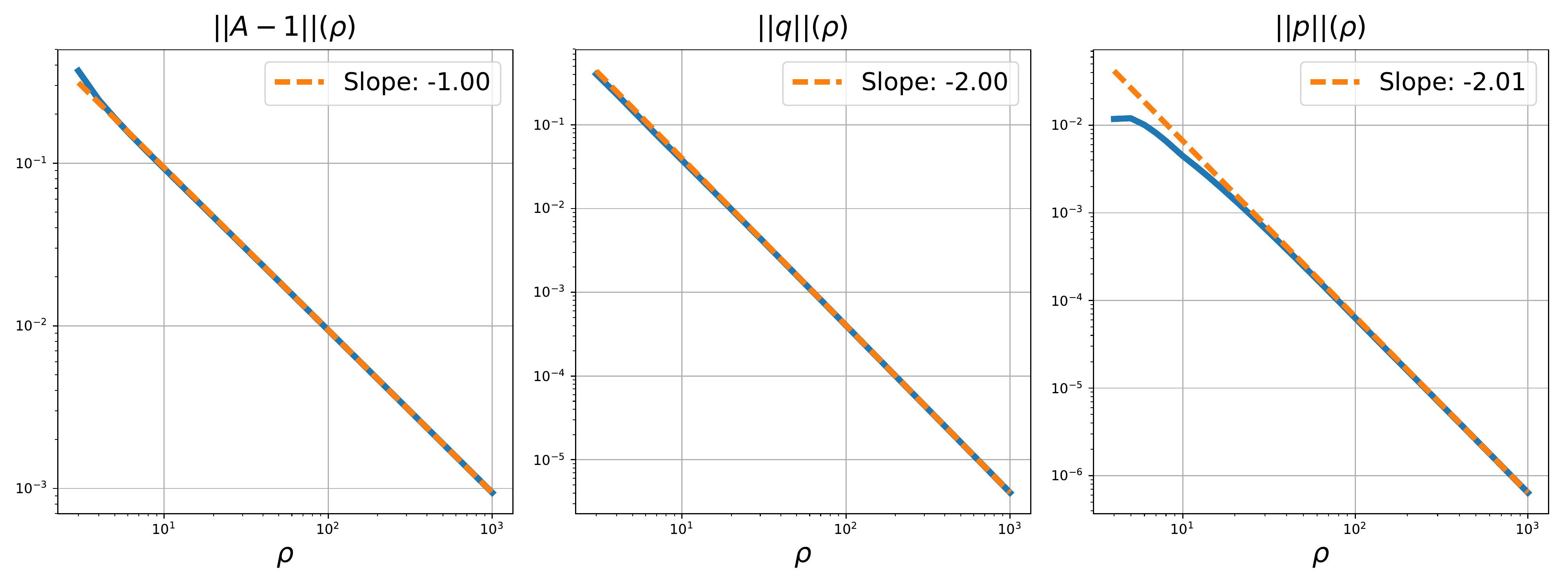}
	 	\caption{Decay plots of the numerical solution for the ``binary black hole case'' obtained with $M_+=M_-=1/2$, $Z=1$, $N=11$, $\rho_0=3$ and a numerical error tolerance of $10^{-12}$.} 
	 	\label{fig:decaynormsbbh}
	 \end{figure}
         \begin{figure}[t]
           \begin{minipage}{0.49\linewidth}
		\centering
		\includegraphics[width=0.85\linewidth]{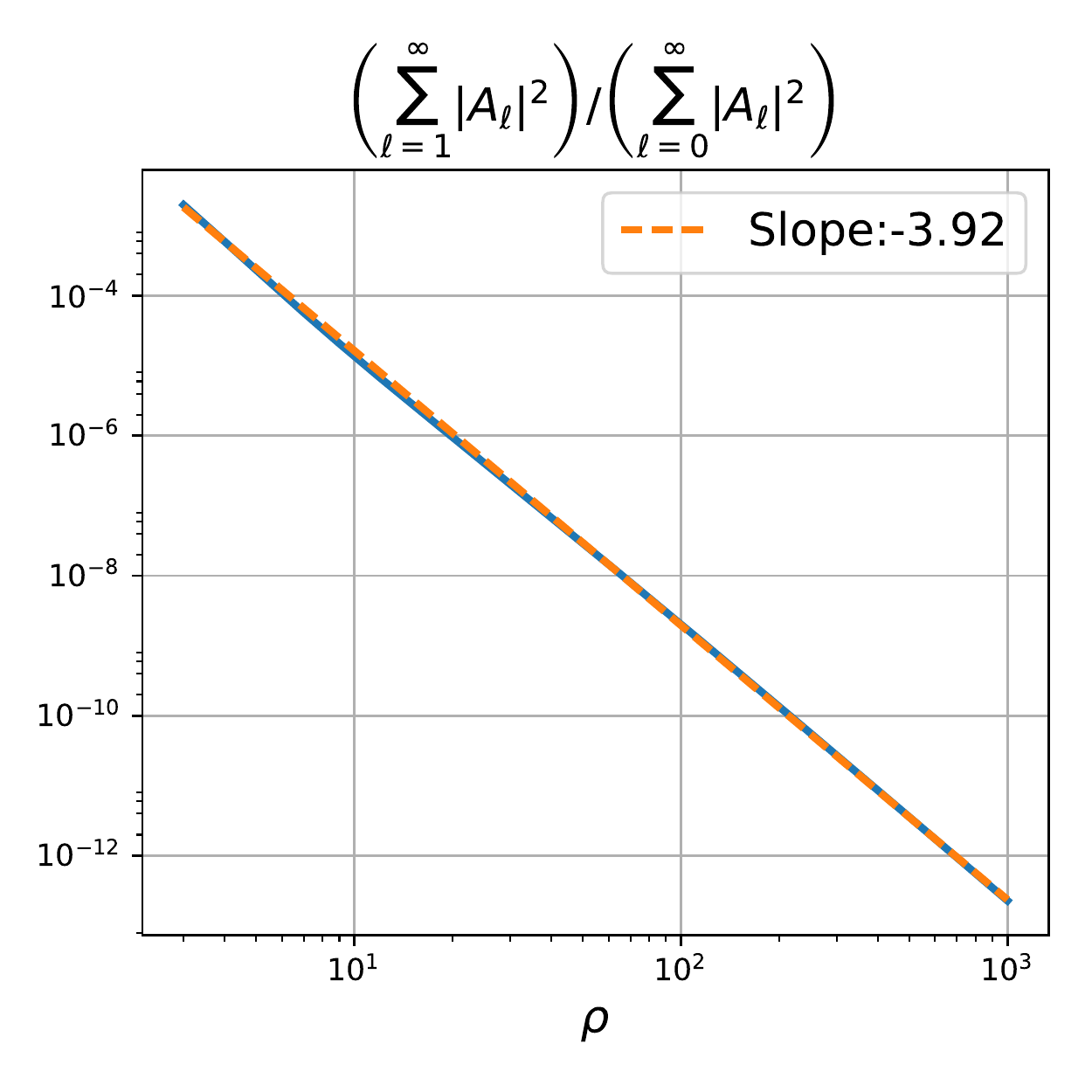}
		\caption{Mode decay plot of the numerical solution for the ``binary black hole case'' obtained with the same parameters as \Figref{fig:decaynormsbbh}.}
		\label{fig:massdecaybbh}
              \end{minipage}%
              \hfill%
              \begin{minipage}{0.49\linewidth}
                \centering
                \includegraphics[width=\linewidth]{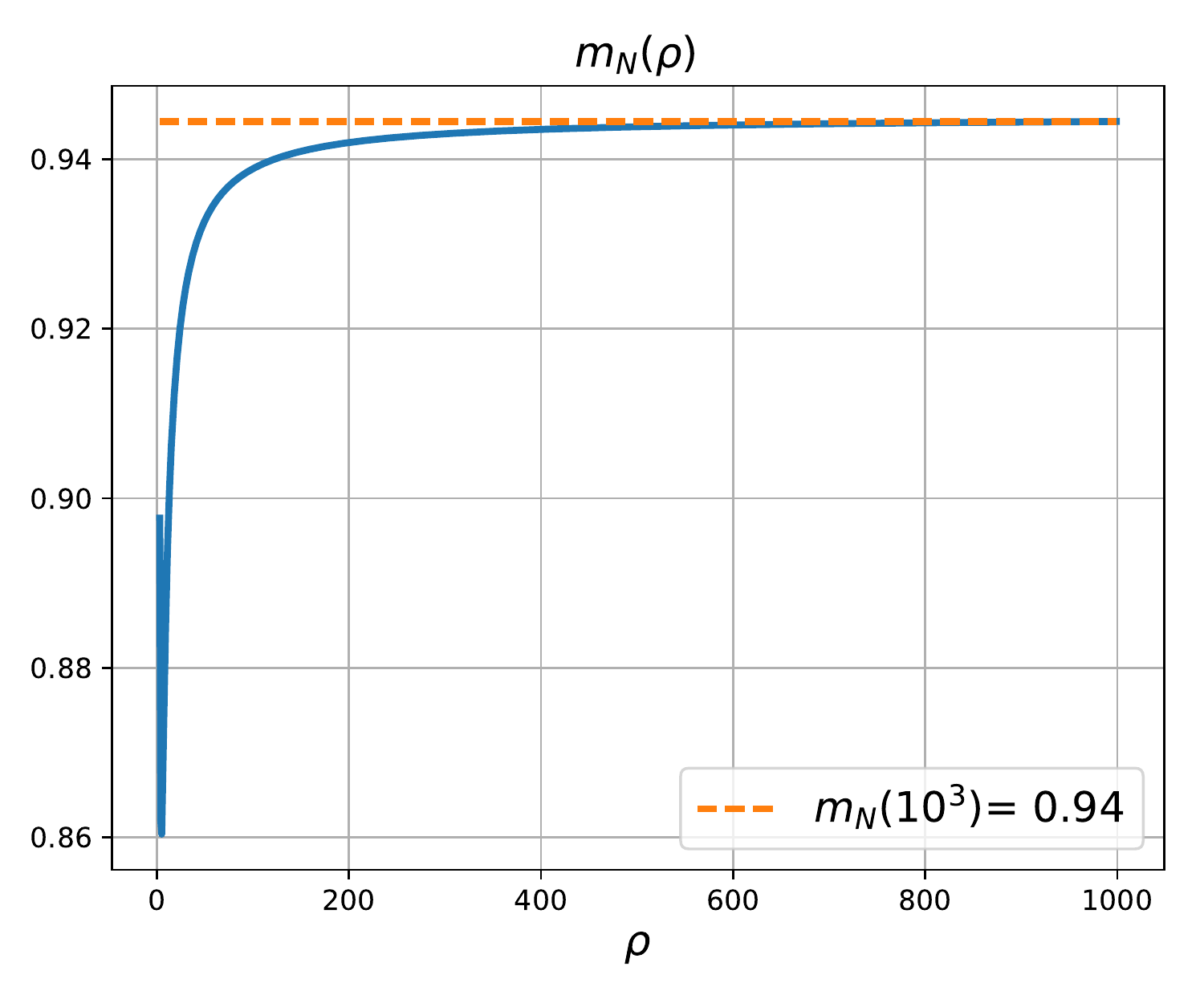}
     	\caption{Estimate of the ADM mass for the ``binary black hole case''  obtained with the same parameters as \Figref{fig:decaynormsbbh}.}
     	\label{fig:massplotbbh}
      \end{minipage}
    \end{figure}

        Our numerical findings, as shown in \Figref{fig:decaynormsbbh}, are again consistent with the prediction 
	\begin{align}
	\| A -1 \|(\rho)= \DecayO{\rho}{},\quad \| q \|(\rho) = \DecayO{\rho}{2},\quad \| p\|(\rho) = \DecayO{\rho}{2}
	\end{align}
     from \Resultref{Result:AsymAnal_Flt_A}.
          Similarly, as with the single black hole case, we expect the quantity
     \begin{align}
     \frac{\sum_{\ell=1}^{\infty}|A_\ell (\rho)|^2 }{ \sum_{\ell=0}^{\infty}|A_\ell (\rho)|^2 }
     \end{align}
     to decay like $O(\rho^{-4})$. In \Figref{fig:massdecaybbh} we observe exactly this behaviour. As before, we interpret this as strong evidence that the obtained vacuum initial data sets are indeed asymptotically flat.
     One may therefore use \Eqref{Eq:ADM_Numerical} to numerically estimate the ADM mass ; the behaviour predicted by \Eqref{eq:admmassconvergence} is verified in \Figref{fig:massplotbbh}. We find          
     \begin{align}
     m_{ADM}=0.9423,\;\;\; \mathcal{E}_{A}\left[m_{ADM}\right]= 5.01\times 10^{-6}.
     \end{align}    

	\begin{figure}[t]
		\centering
		\includegraphics[width=0.40\linewidth]{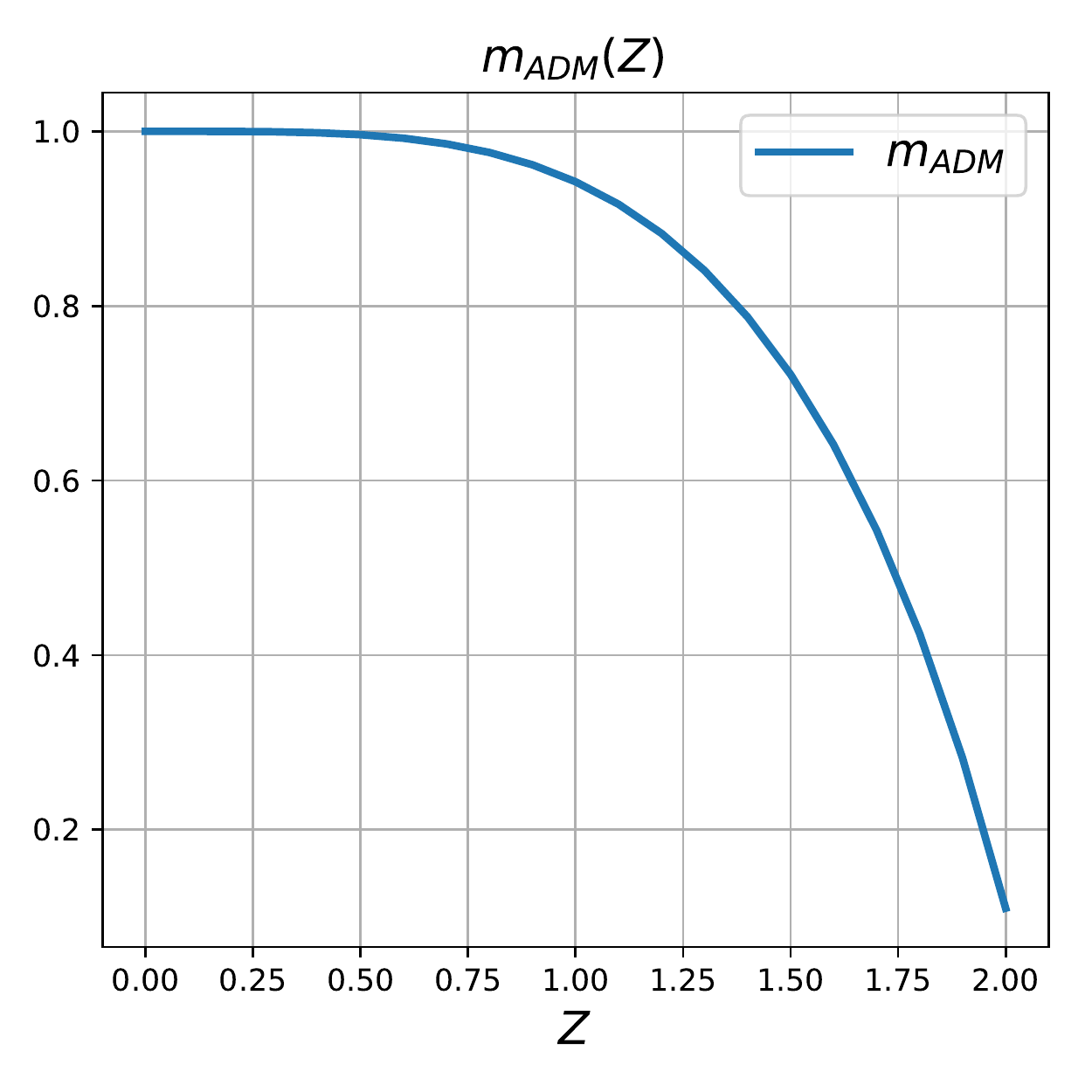}
		\caption{Dependence of $m_{ADM}$ on $Z$ in the ``binary black hole case'' with $M_{+}=M_{-}=1/2$, $\rho_0=3$, $N=11$ and numerical error tolerance of $10^{-12}$. }
		\label{fig:massbbh}
	\end{figure}
	We have repeated the calculations for similar parameter sets and came to the same conclusions: The resulting vacuum initial data sets are always asymptotically flat.
	Given fixed values of $M_+$ and $M_-$, say, $M_+=M_-=1/2$ as before, one expects the resulting ADM masses to depend strongly on the separation distance $Z$. To investigate this we numerically calculate the resulting vacuum initial data sets and  ADM masses for a range of separation distances $Z$. Note that since we treat $\rho_{0}=3$ as fixed, \Eqref{Eq:BirfurificationPoint} introduces an upper bound for the possible values for $Z$, namely $Z<\rho_{crit}$. The results are shown in \Figref{fig:massbbh}, where we see that the ADM mass is a decreasing function of the separation distance $Z$. Notice that the same dependence of the ADM-mass on $Z$ had been observed in \cite{Beyer:2018HW} for asymptotically Euclidean data sets.

	\section{Conclusions}
	In this paper we propose a new parabolic-hyperbolic formulation of the Einstein vacuum constraints based on a formulation originally given by R\'{a}cz. Using analytical and numerical methods we provide strong evidence that the main major drawback of these kinds of evolutionary formulations, namely to generically produce vacuum initial data sets which violate asymptotically flatness \cite{Beyer:2017tu,Beyer:2018HW,csukas2019}, has now finally been overcome.

In Sections~\ref{SubSec:Perturbing_spherically_symmetric_initial_data_sets} and \ref{SubSec:BBH_initial_data_sets} we have numerically constructed particular vacuum initial data sets as solutions of our new equations which could potentially be interpreted as  perturbed Schwarzschild initial data and as binary black hole initial data, respectively. As we discussed, the particular choice of foliation (see \Sectionref{Sec:A_superposition_method_for_generating_binary_black_hole_data}) leads to the restriction $\rho>\rho_{crit}$ with \Eqref{Eq:BirfurificationPoint}. This means that we only have limited access to the strong field regime close to the  black holes. Strictly speaking it is therefore not even clear whether the resulting vacuum initial data sets really represent black holes. In order to resolve this issue, we need to find for example  apparent horizons in the strong field regime. Given that the asymptotics of the resulting initial data sets are under control now, future studies will therefore have to focus on a remedy for the issues associated with the strong field regime. A natural starting point for such studies would be to try to come up with a different $2$-sphere foliation than the one in  \Sectionref{Sec:A_superposition_method_for_generating_binary_black_hole_data}, which matches the one above for sufficiently large values of $\rho$, but which allows to place the \emph{initial $2$-surface arbitrarily close to the black holes}. All this would need to be done in a way which guarantees that $\kstar$ is strictly negative, which might be a non-trivial condition given how involved and non-trivial typical strong field geometries can be. In any case, if this can be achieved, then we can use \ParabolicHyperbolicA to construct asymptotically flat vacuum initial data sets and analyse in great detail the resulting strong field black hole-like regimes.

	\section*{Acknowledgements}
	
	JR is supported by a Ph.D scholarship awarded by the University of Otago. Part of this research was funded by a grant to JF from the Division of Sciences of the University of Otago.
	
	\begin{appendices}
		\setcounter{equation}{0} 
		\numberwithin{equation}{section} 
		\appendixpage
		\appendix
		
		\section{Spin-weight and spin-weighted spherical harmonics}\label{Sec:SWSHstuff}
		We say that a function $f$ defined on $\mathbb{S}^2$ has \emph{spin-weight $s$} if it transforms as $f \to e^{\text{i} s \xi} f$ under a local rotation by an angle $\xi$ in the tangent plane at any point in $\mathbb{S}^2$. Let $(\vartheta,\varphi)$ be standard polar coordinates on $\mathbb S^2$. If $f$ has spin-weight $s$ and is sufficiently smooth, it can be written as
		\begin{equation}\label{eq:functionS2}
		f(\vartheta,\varphi)=  \sum\limits_{l=|s|}^{\infty}  \sum\limits_{m=-l}^{l} f_{lm}\, {}_{s}Y_{lm} (\vartheta,\varphi),
		\end{equation}
		where $_{s}Y_{lm}( \vartheta , \varphi)$ are the \emph{spin-weighted spherical harmonics (SWSH)} and where $f_{lm}$ are complex numbers. Using the conventions in \cite{Penrose:1984tf,Beyer:2015bv,Beyer:2014bu,Beyer:2016fc,Beyer:2017jw,Beyer:2017tu}, these functions satisfy
		\begin{equation}\label{integral_properties_spherical_harmonics}
		\int \limits_{\mathbb{S}^2} \  {}_{s} Y_{l_1 m_1 }(\vartheta,\varphi)
		\: _{s}\overline{Y}_{l_2 m_2}(\vartheta,\varphi) \ d\Omega = \delta_{l_1 l_2} \delta_{m_1 m_2},
		\end{equation}
		where $\delta_{lm}$ is the Kronecker delta and $d\Omega$ is the area element of the metric of the round unit sphere. Using this we find that the coefficients $f_{lm}$ in \Eqref{eq:functionS2} can be calculated as
		\begin{equation}
		f_{lm}=\int \limits_{\mathbb{S}^2} f(\vartheta,\varphi)\, {}_{s}\overline{Y}_{lm} (\vartheta,\varphi) d\Omega.
		\end{equation}
		
		The \textit{eth-operators} $\eth$ and $\eth'$ are  defined by 
		\begin{equation}\label{eq:def_eths}
		\eth f       = \partial_\vartheta f - \dfrac{\text{i}}{ \sin \vartheta} \partial_\varphi f- s f \cot \vartheta, \quad 
		\eth' f = \partial_\vartheta f + \dfrac{\text{i}}{ \sin \vartheta} \partial_\varphi f + s f \cot \vartheta  ,
		\end{equation}
		for any function $f$ on $\mathbb{S}^2$ with spin-weight $s$. We have
		\begin{align}\label{eq:eths}
		\eth  \hspace{0.1cm}_{s}Y_{lm} (\vartheta,\varphi)  &= - \sqrt{ (l-s)(l+s+1) } \hspace{0.1cm}_{s+1}Y_{lm} (\vartheta,\varphi) , \\
		\label{eq:eths2}
		\eth'   \hspace{0.1cm}_{s}Y_{lm} (\vartheta,\varphi)   &= \sqrt{ (l+s)(l-s+1) } \hspace{0.1cm}_{s-1}Y_{lm} (\vartheta,\varphi) , \\
		\eth' \eth  \hspace{0.1cm}_{s}Y_{lm} (\vartheta,\varphi)   &= - (l-s)(l+s+1) \hspace{0.1cm}_{s}Y_{lm} (\vartheta,\varphi) .
		\end{align}
		Thus, using the properties above it is easy to see that $\eth$ raises the spin-weight by one while $\eth'$ lowers it by one.
		
		The \emph{average} of a function $f$ with spin-weight $0$ on $\mathbb{S}^2$ is defined by 
		\begin{equation}
		\label{eq:average}
		\underline{f} = \dfrac{1}{4 \pi} \int \limits_{\mathbb{S}^2} \: f d\Omega.
		\end{equation}
		Expressing $f$ in terms of SWSH and using \Eqref{integral_properties_spherical_harmonics} it follows 
		\begin{equation}\label{ec:mean_value_s2}
		\begin{aligned}
		\underline{f} &= \dfrac{1}{4 \pi}  \int \limits_{\mathbb{S}^2} \: \sum\limits_{l=0}^{\infty}  \sum\limits_{m=-l}^{l} f_{lm}\, {}_{0}Y_{lm} (\vartheta,\varphi) \; d\Omega , \\
		&= \dfrac{\sqrt{4\pi}}{4 \pi }  \int \limits_{\mathbb{S}^2} \: \sum\limits_{l=0}^{\infty}  \sum\limits_{m=-l}^{l} f_{lm}\,{}_{0}Y_{lm} (\vartheta,\varphi) \; _{0}\overline{Y}_{00}(\vartheta,\varphi) \; d\Omega , \\
		& =   \frac{1}{\sqrt{4\pi}}f_{00},
		\end{aligned}
		\end{equation}
		where we have used the fact that $_{0}Y_{00}(\vartheta,\varphi) =  (4\pi)^{-1/2}$. 
		Another quantity of interest is the $L^2$-norm with respect to the standard round metric on $S^2$. The \emph{Parseval identity} states that
		\begin{equation}
		\label{eq:L2}
		\|f\|^2_{L^2(\mathbb{S}^2)}=\sum_{l=0}^\infty\sum_{m=-l}^l |f_{lm}|^2.
		\end{equation}
		
		Finally we notice that many quantities considered in this paper are axially symmetric and therefore do not depend on the angle $\varphi$. For such functions, all coefficients with $f_{lm}$ with $m\not=0$ vanish and we use the following short-hand notation to write \Eqref{eq:functionS2} as
		\begin{equation}\label{eq:functionS2axial}
		f(\vartheta)=  \sum\limits_{l=|s|}^{\infty}  f_{l}\, {}_{s}Y_{l} (\vartheta).
		\end{equation}
		
	\end{appendices}

\begin{thebibliography}{10}

\bibitem{Alcubierre:Book}
M.~Alcubierre.
\newblock {\em Introduction to 3+1 {{Numerical Relativity}}}.
\newblock {Oxford Science Publications}, 2008.

\bibitem{anderson2018a}
M.~T. Anderson.
\newblock On the conformal method for the {{Einstein}} constraint equations.
\newblock 2018.
\newblock Preprint. \href{http://arxiv.org/abs/1812.06320}{arXiv:1812.06320}.

\bibitem{bartnik1993}
R.~A. Bartnik.
\newblock Quasi-spherical metrics and prescribed scalar curvature.
\newblock {\em J. Diff. Geom.}, 37(1):31--71, 1993.
\newblock
  DOI:~\href{https://doi.org/10.4310/jdg/1214453422}{10.4310/jdg/1214453422}.

\bibitem{bartnik2004}
R.~A. Bartnik and J.~Isenberg.
\newblock The {{Constraint Equations}}.
\newblock In {\em The {{Einstein Equations}} and the {{Large Scale Behavior}}
  of {{Gravitational Fields}}}, pages 1--38. {Birkh{\"a}user}, {Basel}, 2004.

\bibitem{Baumgarte:2010vs}
T.~W. Baumgarte and S.~L. Shapiro.
\newblock {\em Numerical {{Relativity}}}.
\newblock Solving {{Einstein}}'s {{Equations}} on the {{Computer}}. {Cambridge
  University Press}, 2010.

\bibitem{Beyer:2009vw}
F.~Beyer.
\newblock A spectral solver for evolution problems with spatial
  {$S^3$}-topology.
\newblock {\em J. Comp. Phys.}, 228(17):6496--6513, 2009.
\newblock
  DOI:~\href{https://doi.org/10.1016/j.jcp.2009.05.037}{10.1016/j.jcp.2009.05.037}.

\bibitem{Beyer:2015bv}
F.~Beyer, B.~Daszuta, and J.~Frauendiener.
\newblock A spectral method for half-integer spin fields based on spin-weighted
  spherical harmonics.
\newblock {\em Class. Quantum Grav.}, 32(17):175013, 2015.
\newblock
  DOI:~\href{https://doi.org/10.1088/0264-9381/32/17/175013}{10.1088/0264-9381/32/17/175013}.

\bibitem{Beyer:2014bu}
F.~Beyer, B.~Daszuta, J.~Frauendiener, and B.~Whale.
\newblock Numerical evolutions of fields on the 2-sphere using a spectral
  method based on spin-weighted spherical harmonics.
\newblock {\em Class. Quantum Grav.}, 31(7):075019, 2014.
\newblock
  DOI:~\href{https://doi.org/10.1088/0264-9381/31/7/075019}{10.1088/0264-9381/31/7/075019}.

\bibitem{Beyer:2016fc}
F.~Beyer, L.~Escobar, and J.~Frauendiener.
\newblock Numerical solutions of {{Einstein}}'s equations for cosmological
  spacetimes with spatial topology {$S^3$} and symmetry group {$U(1)$}.
\newblock {\em Phys. Rev. D}, 93(4):043009, 2016.
\newblock
  DOI:~\href{https://doi.org/10.1103/PhysRevD.93.043009}{10.1103/PhysRevD.93.043009}.

\bibitem{Beyer:2017tu}
F.~Beyer, L.~Escobar, and J.~Frauendiener.
\newblock Asymptotics of solutions of a hyperbolic formulation of the
  constraint equations.
\newblock {\em Class. Quantum Grav.}, 34(20):205014, 2017.
\newblock
  DOI:~\href{https://doi.org/10.1088/1361-6382/aa8be6}{10.1088/1361-6382/aa8be6}.

\bibitem{Beyer:2017jw}
F.~Beyer, L.~Escobar, and J.~Frauendiener.
\newblock Criticality of inhomogeneous {{Nariai}}-like cosmological models.
\newblock {\em Phys. Rev. D}, 95(8):084030, 2017.
\newblock
  DOI:~\href{https://doi.org/10.1103/PhysRevD.95.084030}{10.1103/PhysRevD.95.084030}.

\bibitem{Beyer:2018HW}
F.~Beyer, L.~Escobar, J.~Frauendiener, and J.~Ritchie.
\newblock Numerical construction of initial data sets of binary black hole type
  using a parabolic-hyperbolic formulation of the vacuum constraint equations.
\newblock {\em Class. Quantum Grav.}, 36(17):175005, 2019.
\newblock
  DOI:~\href{https://doi.org/10.1088/1361-6382/ab3482}{10.1088/1361-6382/ab3482}.

\bibitem{Bishop:2004gb}
N.~T. Bishop, F.~Beyer, and M.~Koppitz.
\newblock Black hole initial data from a nonconformal decomposition.
\newblock {\em Phys. Rev. D}, 69(6):064010, 2004.
\newblock
  DOI:~\href{https://doi.org/10.1103/PhysRevD.69.064010}{10.1103/PhysRevD.69.064010}.

\bibitem{Bishop:1998cb}
N.~T. Bishop, R.~Isaacson, M.~Maharaj, and J.~Winicour.
\newblock Black hole data via a {{Kerr}}-{{Schild}} approach.
\newblock {\em Phys. Rev. D}, 57(10):6113--6118, 1998.
\newblock
  DOI:~\href{https://doi.org/10.1103/PhysRevD.57.6113}{10.1103/PhysRevD.57.6113}.

\bibitem{Cerebaum:2016}
C.~Cederbaum, J.~Cortier, and A.~Sakovich.
\newblock On the {{Center}} of {{Mass}} of {{Asymptotically Hyperbolic Initial
  Data Sets}}.
\newblock {\em Ann. Henri Poincar{\'e}}, 17(6):1505--1528, 2016.
\newblock
  DOI:~\href{https://doi.org/10.1007/s00023-015-0438-5}{10.1007/s00023-015-0438-5}.

\bibitem{ChoquetBruhat:1969cl}
Y.~{Choquet-Bruhat} and R.~P. Geroch.
\newblock Global aspects of the {{Cauchy}} problem in general relativity.
\newblock {\em Commun. Math. Phys.}, 14(4):329--335, 1969.
\newblock DOI:~\href{https://doi.org/10.1007/BF01645389}{10.1007/BF01645389}.

\bibitem{choquet-bruhat2000}
Y.~{Choquet-Bruhat}, J.~Isenberg, and J.~W. York.
\newblock Einstein constraints on asymptotically {{Euclidean}} manifolds.
\newblock {\em Phys. Rev. D}, 61(8), 2000.
\newblock
  DOI:~\href{https://doi.org/10.1103/PhysRevD.61.084034}{10.1103/PhysRevD.61.084034}.

\bibitem{chu2014}
T.~Chu.
\newblock Including realistic tidal deformations in binary black-hole initial
  data.
\newblock {\em Phys. Rev. D}, 89(6):064062, 2014.
\newblock
  DOI:~\href{https://doi.org/10.1103/PhysRevD.89.064062}{10.1103/PhysRevD.89.064062}.

\bibitem{csukas2019}
K.~Csuk{\'a}s and I.~R{\'a}cz.
\newblock On the asymptotics of solutions to the evolutionary form of the
  constraints.
\newblock 2019.
\newblock Preprint. \href{http://arxiv.org/abs/1911.02900}{arXiv:1911.02900}.

\bibitem{Dain:2001cd}
S.~Dain and H.~Friedrich.
\newblock Asymptotically {{Flat Initial Data}} with {{Prescribed Regularity}}
  at {{Infinity}}.
\newblock {\em Commun. Math. Phys.}, 222(3):569--609, 2001.
\newblock
  DOI:~\href{https://doi.org/10.1007/s002200100524}{10.1007/s002200100524}.

\bibitem{dilts2017}
J.~Dilts, M.~Holst, T.~Kozareva, and D.~Maxwell.
\newblock Numerical {{Bifurcation Analysis}} of the {{Conformal Method}}.
\newblock 2017.
\newblock Preprint. \href{http://arxiv.org/abs/1710.03201}{arXiv:1710.03201}.

\bibitem{doulis2019}
G.~Doulis.
\newblock Construction of high precision numerical single and binary black hole
  initial data.
\newblock {\em Phys. Rev. D}, 100(2):024064, 2019.
\newblock
  DOI:~\href{https://doi.org/10.1103/PhysRevD.100.024064}{10.1103/PhysRevD.100.024064}.

\bibitem{FouresBruhat:1952ji}
Y.~{Four{\`e}s-Bruhat}.
\newblock Th{\'e}or{\`e}me d'existence pour certains syst{\`e}mes
  d'{\'e}quations aux d{\'e}riv{\'e}es partielles non lin{\'e}aires.
\newblock {\em Acta Math.}, 88(1):141--225, 1952.
\newblock DOI:~\href{https://doi.org/10.1007/BF02392131}{10.1007/BF02392131}.

\bibitem{hawking1968}
S.~W. Hawking.
\newblock Gravitational {{Radiation}} in an {{Expanding Universe}}.
\newblock {\em J. Math. Phys.}, 9(4):598--604, 1968.
\newblock DOI:~\href{https://doi.org/10.1063/1.1664615}{10.1063/1.1664615}.

\bibitem{lovelace2009}
G.~Lovelace.
\newblock Reducing spurious gravitational radiation in binary-black-hole
  simulations by using conformally curved initial data.
\newblock {\em Class. Quantum Grav.}, 26(11):114002, 2009.
\newblock
  DOI:~\href{https://doi.org/10.1088/0264-9381/26/11/114002}{10.1088/0264-9381/26/11/114002}.

\bibitem{Matzner:1998hv}
R.~A. Matzner, M.~F. Huq, and D.~Shoemaker.
\newblock Initial data and coordinates for multiple black hole systems.
\newblock {\em Phys. Rev. D}, 59(2):024015, 1998.
\newblock
  DOI:~\href{https://doi.org/10.1103/PhysRevD.59.024015}{10.1103/PhysRevD.59.024015}.

\bibitem{Moreno:2002dm}
C.~Moreno, D.~N{\'u}{\~n}ez, and O.~Sarbach.
\newblock Kerr\textendash{{Schild}}-type initial data for black holes with
  angular momenta.
\newblock {\em Class. Quantum Grav.}, 19(23):6059--6073, 2002.
\newblock
  DOI:~\href{https://doi.org/10.1088/0264-9381/19/23/312}{10.1088/0264-9381/19/23/312}.

\bibitem{Nakonieczna:2017vk}
A.~Nakonieczna, {\L}.~Nakonieczny, and I.~R{\'a}cz.
\newblock Black hole initial data by numerical integration of the
  parabolic-hyperbolic form of the constraints.
\newblock 2017.
\newblock Preprint. \href{http://arxiv.org/abs/1712.00607}{arXiv:1712.00607}.

\bibitem{Penrose:1984tf}
R.~Penrose and W.~Rindler.
\newblock {\em Two-{{Spinor Calculus}} and {{Relativistic Fields}}}, volume~1
  of {\em Spinors and {{Space}}-{{Time}}}.
\newblock {Cambridge University Press}, {Cambridge}, 1984.

\bibitem{Racz:2014dx}
I.~R{\'a}cz.
\newblock Is the {{Bianchi}} identity always hyperbolic?
\newblock {\em Class. Quantum Grav.}, 31(15):155004, 2014.
\newblock
  DOI:~\href{https://doi.org/10.1088/0264-9381/31/15/155004}{10.1088/0264-9381/31/15/155004}.

\bibitem{Racz:2014kk}
I.~R{\'a}cz.
\newblock Cauchy problem as a two-surface based `geometrodynamics'.
\newblock {\em Class. Quantum Grav.}, 32(1):015006, 2015.
\newblock
  DOI:~\href{https://doi.org/10.1088/0264-9381/32/1/015006}{10.1088/0264-9381/32/1/015006}.

\bibitem{Racz:2015gb}
I.~R{\'a}cz.
\newblock Constraints as evolutionary systems.
\newblock {\em Class. Quantum Grav.}, 33(1):015014, 2016.
\newblock
  DOI:~\href{https://doi.org/10.1088/0264-9381/33/1/015014}{10.1088/0264-9381/33/1/015014}.

\bibitem{racz2018}
I.~R{\'a}cz.
\newblock On the {{Evolutionary Form}} of the {{Constraints}} in
  {{Electrodynamics}}.
\newblock {\em Symmetry}, 11(1):10, 2018.
\newblock DOI:~\href{https://doi.org/10.3390/sym11010010}{10.3390/sym11010010}.

\bibitem{Racz:2015bu}
I.~R{\'a}cz and J.~Winicour.
\newblock Black hole initial data without elliptic equations.
\newblock {\em Phys. Rev. D}, 91(12):124013, 2015.
\newblock
  DOI:~\href{https://doi.org/10.1103/PhysRevD.91.124013}{10.1103/PhysRevD.91.124013}.

\bibitem{winicourRacz2018}
I.~R{\'a}cz and J.~Winicour.
\newblock Toward computing gravitational initial data without elliptic solvers.
\newblock {\em Class. Quantum Grav.}, 35(13):135002, 2018.
\newblock
  DOI:~\href{https://doi.org/10.1088/1361-6382/aac5c5}{10.1088/1361-6382/aac5c5}.

\bibitem{Szabados:2009ig}
L.~B. Szabados.
\newblock Quasi-{{Local Energy}}-{{Momentum}} and {{Angular Momentum}} in
  {{General Relativity}}.
\newblock {\em Living Rev. Relativity}, 12(4):4, 2009.
\newblock DOI:~\href{https://doi.org/10.12942/lrr-2009-4}{10.12942/lrr-2009-4}.

\bibitem{Winicour:2017wr}
J.~Winicour.
\newblock The algebraic-hyperbolic approach to the linearized gravitational
  constraints on a {{Minkowski}} background.
\newblock {\em Class. Quantum Grav.}, 34(15):157001, 2017.
\newblock
  DOI:~\href{https://doi.org/10.1088/1361-6382/aa7bd6}{10.1088/1361-6382/aa7bd6}.

\end{thebibliography}

\end{document}